\documentclass[twocolumn,aps,epsfig,nofootinbib]{revtex4}
\usepackage{graphicx}
\usepackage{dcolumn}
\usepackage{amsmath}
\usepackage{amssymb}
\usepackage{epsfig}
\usepackage[colorlinks,linkcolor=blue,citecolor=blue,urlcolor=blue]{hyperref}

\begin{document}

\title{Pre-inflationary dynamics in loop quantum cosmology: Power-law potentials }
\author{M. Shahalam$^1$ \footnote{E-mail address: shahalam@zjut.edu.cn}}
\author{Manabendra Sharma$^1$ \footnote{ E-mail address: manabendra@zjut.edu.cn}}
\author{Qiang Wu$^1$ \footnote{E-mail address: wuq@zjut.edu.cn}}
\author{Anzhong Wang$^{1,2}$ \footnote{E-mail address: Anzhong$\_$Wang@baylor.edu}}
\affiliation{$^{1}$Institute for Advanced Physics $\&$ Mathematics,
Zhejiang University of Technology, Hangzhou, 310032, China\\
$^2$GCAP-CASPER, Department of Physics, Baylor University, Waco, TX, 76798-7316, USA}

\date{\today}

\begin{abstract}
In this paper, we study the pre-inflationary dynamics for the power-law potential $(V(\phi) \propto \phi^n)$ with $n<2$ in the framework 
of loop quantum cosmology. In the case where the kinetic energy of the inflaton dominates at the initial, the evolution of the universe 
can always be divided into three different phases prior to reheating: {\em  bouncing, transition and slow-roll inflation}. During the 
bouncing phase, the evolution of the expansion factor is independent not only on the initial conditions but also the inflationary potentials, 
and is given explicitly by an analytical solution. In contrast, for the potential energy dominated initial conditions, this universality is lost. 
We also obtain total number of e-folds during the slow-roll inflation, whereby physically viable models are identified. In addition, we present  
phase space analysis for the inflationary potentials under consideration and compare our results with the ones obtained previously for 
different potentials.

\end{abstract}
\maketitle

\section{Introduction}
\label{intro}
The cosmic inflation resolves many problems in the standard model of cosmology such as the horizon and flatness problems, etc. Inflation also describes the formation of the large scale structure of the universe and the origin of inhomogeneities observed in the cosmic microwave background (CMB) \cite{guth1981}. There are large number of inflationary models that can be compatible with the observations. Recently, Planck 2015 results show that, in the case of a single field inflation, the simple quadratic potential model with $(V(\phi) \propto \phi^n$ with $n=2)$ is moderately disfavored in comparison to the power-law with $n<2$ and Starobinsky potentials \cite{Planck2015}. To this effect, it increases the curiosity in studying a single field inflation for the power-law potential with $n<2$.

Despite the huge success of the standard inflationary models that are based on classical general relativity (GR), its past is insufficient due to the existence of a big bang singularity. It occurs in all scalar field models of inflation. To explain the inflationary process in context of GR, the initial singularity is inevitable \cite{borde1994,borde2003}, with which it is not very clear when and how to set the initial conditions. Additionally, to be compatible with the present observations, the universe has to expand at least 60 e-folds during the inflation. Nevertheless, there are more than 70 e-folds during inflation in a large class of inflationary models \cite{martin2014}. In such models, the size of the present universe is less than that of the Planck at the beginning of inflation. As a result, the usual semi-classical treatments during inflation are questionable. This is known as the trans-Planckian problem \cite{martin2001,berger2013}. In addition, one normally ignores the pre-inflationary dynamics and sets the Bunch-Davies (BD) vacuum state at the time when the wavelength of fluctuations were inside the Hubble horizon during inflation. However, the pre-inflationary dynamics can give rise to non-BD states at the onset of inflation \cite{Tao2017}. 

To address the above important issues, one way is to work in the framework of  loop quantum cosmology (LQC), which provides a viable  description of inflation together with its pre-inflationary dynamics \cite{Tao2017,agullo2013a,agullo2013b,agullo2015}. It is interesting  to note that in such a framework the quantum geometrical effects  at Planck scale provide a natural resolution of the big bang singularity \cite{ashtekar2011,ashtekar2015,barrau2016,yang2009}, and the initial singularity is simply replaced by a nonsingular quantum bounce. Even more interesting,  the universe which commences at the bounce can finally lead to the desired slow-roll inflation \cite{ashtekar2010,psingh2006,zhang2007,chen2015,bolliet2015,schander2016,bolliet2016,Bonga2016,Mielczareka}.

To study the pre-inflationary dynamics and cosmological perturbations, there are mainly two different approaches, the dressed metric \cite{agullo2013b,metrica,metricb,metricc} and deformed algebra \cite{algebraa,algebrab,algebrac,algebrad,algebrae,algebraf}. However, as far as only the evolution of the background of the universe is concerned, both approaches provide the same set of dynamical equations for the evolution of the background.  Therefore, the results to be presented in this paper will be applicable to both approaches.  With this in mind,  we shall compare our results with the quadratic and Starobinsky potentials \cite{psingh2006,chen2015,Bonga2016,Tao2017}. In addition, similar analysis can be carried out for other inflationary models, such as, the monodromy, natural inflationary models, and so on. However,  it is expected that the main conclusions obtained in this paper will be valid in those models, too, at least in the case in which the evolution of the universe  is initially dominated by the kinetic energy of the inflaton \cite{Tao2017}. This is because the kinetic energy will dominate the evolution of the universe in the whole bouncing phase, once it dominates it at the initial moment, and the potential energy remains un-dominant until the transition phase, at which the kinetic energy
suddenly drops below the potential energy, whereby the latter takes over, and slow-roll inflation starts \cite{Tao2017}.

The rest of the paper is organized as follows. Section \ref{sec:EOM} is devoted to the detailed analysis of the background evolution in the context of the positive inflaton velocity (PIV, $\dot\phi > 0$) and negative inflaton velocity (NIV, $\dot\phi < 0$), and also kinetic energy dominated (KED) and potential energy dominated (PED) cases at the quantum bounce. In section \ref{sec:phase}, we present the phase space analysis for the power-law potential with $n<2$. Our results are summarized in section \ref{sec:conc}.  

Before proceeding further, we would like to notice that inflation with a power-law potential has been studied in Einstein's theory of gravity \cite{HISY,BG15} and in string-inspired model \cite{SW08}. In particular, it was shown that weak singularity always occurs in high-order derivatives of the scalar field $\left|d^{(k+2)}\phi/d t^{k+2}\right| \rightarrow \infty$ as $\phi \rightarrow 0$ for $k < n < k +1$, where $k$ is an integer \cite{BG15}. In particular, for $0 < n < 1$ the first derivative of the
Ricci scalar diverges, $\dot{R} = 6(4H\dot{H} + \ddot{H}) \rightarrow - \infty$ as $\phi \rightarrow 0$. However, this singularity is weak in the sense that the space-time is geodesically complete and extendible at the singularity \cite{BG15}. Since the Klein-Gordon equation in  both cases are the same, it can be shown that such singularities also occur here in LQC. Moreover, recently inflation for a Bianchi I universe with different inflaton potentials and initial conditions were studied in \cite{killian}, and interesting results were 
obtained.

\section{Equations of motion and Background evolution}
\label{sec:EOM}
In a spatially flat Friedmann-Lemaitre-Robertson-Walker (FLRW) universe, the modified Friedmann equation in the framework of LQC can be written as \cite{ashtekar2006}
\begin{eqnarray}
H^2=\frac{8 \pi}{3 m_{pl}^2}~\rho \Big{(}1-\frac{\rho}{\rho_c}\Big{)}
\label{eq:H}
\end{eqnarray}
where $H=\dot{a}/a$, represents the Hubble parameter, and dot denotes derivative with respect to the cosmic time t, $m_{pl}$ is the Planck mass, $\rho$ denotes the energy density of matter sources and $\rho_c$ is the critical energy density that designates the maximum value of energy density and found to be $\rho_c \simeq 0.41 m_{pl}^4$ \cite{Meissne,Domagala}.

In the framework of LQC, the conservation equation remains the same as in the classical theory,
\begin{eqnarray}
\dot{\rho}+3H(\rho+p)=0
\label{eq:conser}
\end{eqnarray}
where $p$ is the pressure of the matter field. If the matter source is a single scalar field, then the above equation gives the Klein-Gordon equation,
\begin{eqnarray}
\ddot{\phi}+3H \dot{\phi}+ \frac{dV(\phi)}{d\phi}=0
\label{eq:ddphi}
\end{eqnarray}
Equation (\ref{eq:H}) shows that when $\rho=\rho_c$, the Hubble parameter becomes zero which means quantum bounce occurs at $\rho=\rho_c$. In the literature, the background evolution with the bouncing phase has been extensively discussed. One of the important result is that, following bounce, a desired slow-roll inflation is obtained generically \cite{psingh2006,Mielczarek,zhang2007,chen2015,Tao2017,ashtekar2011}. Following this, we shall study ``bounce and slow-roll" with the power-law potentials,  
\begin{eqnarray}
V(\phi)=\frac{1}{2}m^{4-n}\phi^n
\label{eq:pot}
\end{eqnarray}
where $m$ has the dimension of mass. We consider five particular values of $n$: $n = 7/4, 4/3, 1, 2/3$ and $1/3$, respectively, which are all consistent with Planck 2015 results for inflationary universe \cite{Planck2015}. The corresponding values of $m$ for every potential are given as
\begin{eqnarray}
m=6.2 \times 10^{-6}m_{pl}, \qquad n&=&7/4, \nonumber\\
5.1 \times 10^{-5}m_{pl}, \qquad n&=&4/3, \nonumber\\
1.9 \times 10^{-4}m_{pl}, \qquad n&=&1, \nonumber\\
4.7 \times 10^{-4}m_{pl}, \qquad n&=&2/3, \nonumber\\
1.1 \times 10^{-3}m_{pl}, \qquad n&=&1/3.
\label{eq:mass}
\end{eqnarray}
Let us first investigate the evolution equations numerically for different power-law potentials. We solve equations (\ref{eq:H}) and (\ref{eq:ddphi}) numerically with the initial conditions of $a(t)$, $\phi(t)$ and $\dot{\phi}(t)$ at a particular moment. A natural choice 
of the time is at the bounce $t=t_B$,  at which we have
\begin{eqnarray}
\rho_B &=& \rho_c =\frac{1}{2}\dot{\phi}^2(t_B)+V(\phi(t_B)), \nonumber\\
 \dot{a}(t_B)&=&0,
\label{eq:bounce} \end{eqnarray}
which implies that
\begin{eqnarray}
\dot{\phi}(t_B) &=& \pm \sqrt{2 \Big{(} \rho_c - V(\phi(t_B)) \Big{)}}, 
\label{eq:bounce2}
\end{eqnarray}
and a suitable choice for $a(t_B)$ is 
\begin{eqnarray}
a(t_B) &=& 1.
\label{eq:bounce3}
\end{eqnarray}
Here after, we read off $\phi(t_B)$ and $\dot{\phi}(t_B)$ as $\phi_B$ and $\dot{\phi}_B$. Since $\dot{\phi}(t_B)$ is given by Eq.(\ref{eq:bounce2}) for any given potential,   the initial conditions will be uniquely specified by $\phi_B$ only. Following this, we shall consider two cases: (a) PIV: $\dot{\phi}_B > 0$ and (b) NIV: $\dot{\phi}_B < 0$. Let us introduce the following quantities which are important for this paper \cite{Tao2017}.

(1) The equation of state (EOS) $w(\phi)$ for inflaton field is given by
\begin{eqnarray}
w(\phi) = \frac{\dot{\phi}^2/2-V(\phi)}{\dot{\phi}^2/2+V(\phi)}
\label{eq:w}
\end{eqnarray}
The EOS has to be very close to $-1$ during the slow-roll inflation.

(2) The slow-roll parameter $\epsilon_H$, which is defined in terms of $H$ and its derivatives,
\begin{eqnarray}
\epsilon_H = - \frac{\dot{H}}{H^2}
\label{eq:epsilon}
\end{eqnarray}
During the slow-roll inflation, $\epsilon_H \ll 1$.

\begin{figure*}[tbp]
\begin{center}
\begin{tabular}{ccc}
{\includegraphics[width=2.1in,height=1.65in,angle=0]{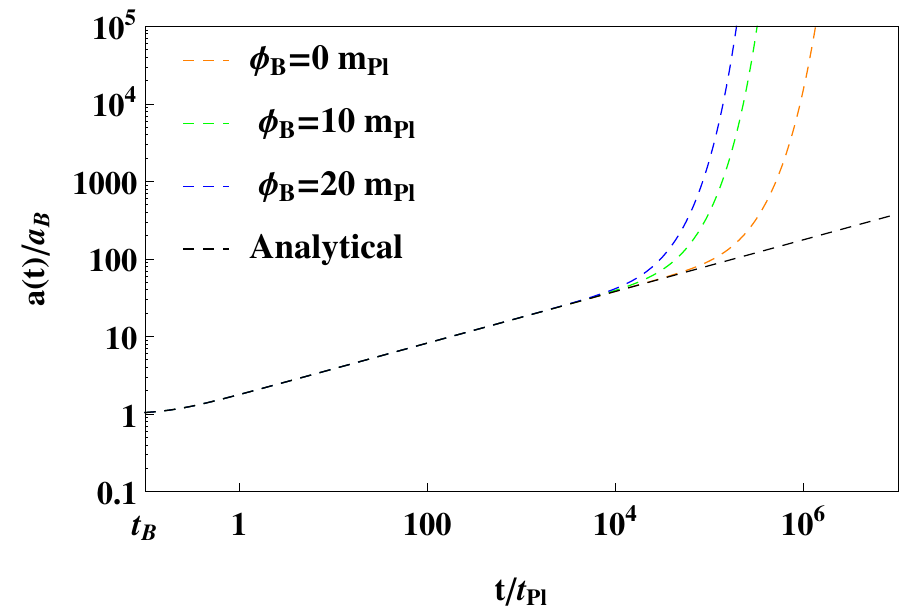}} &
{\includegraphics[width=2.1in,height=1.6in,angle=0]{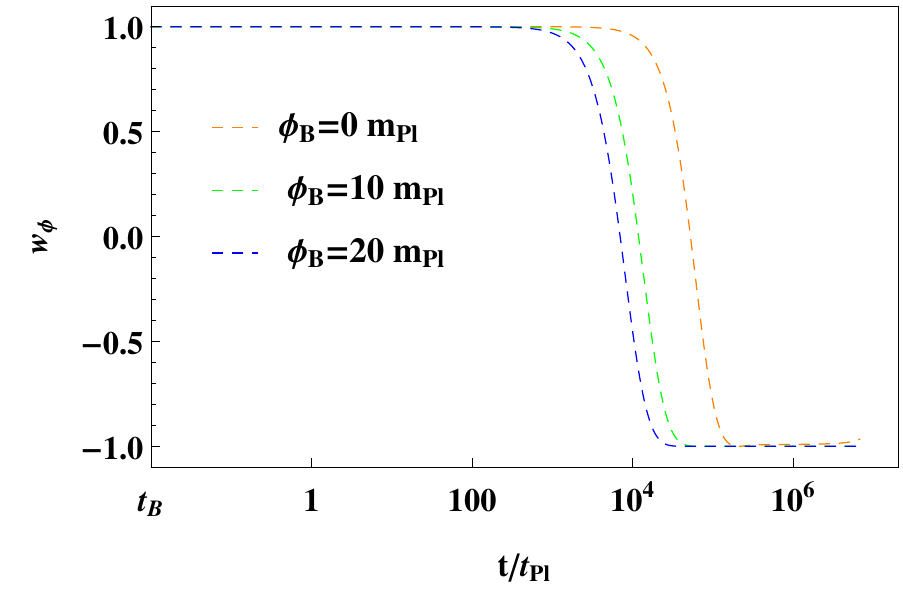}} &
{\includegraphics[width=2.0in,height=1.6in,angle=0]{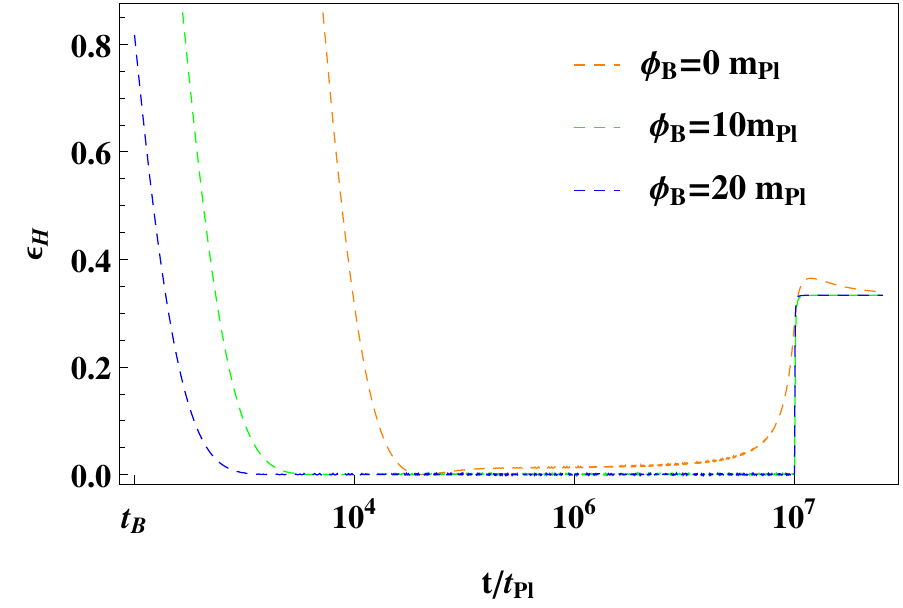}}
 \\
{\includegraphics[width=2.1in,height=1.6in,angle=0]{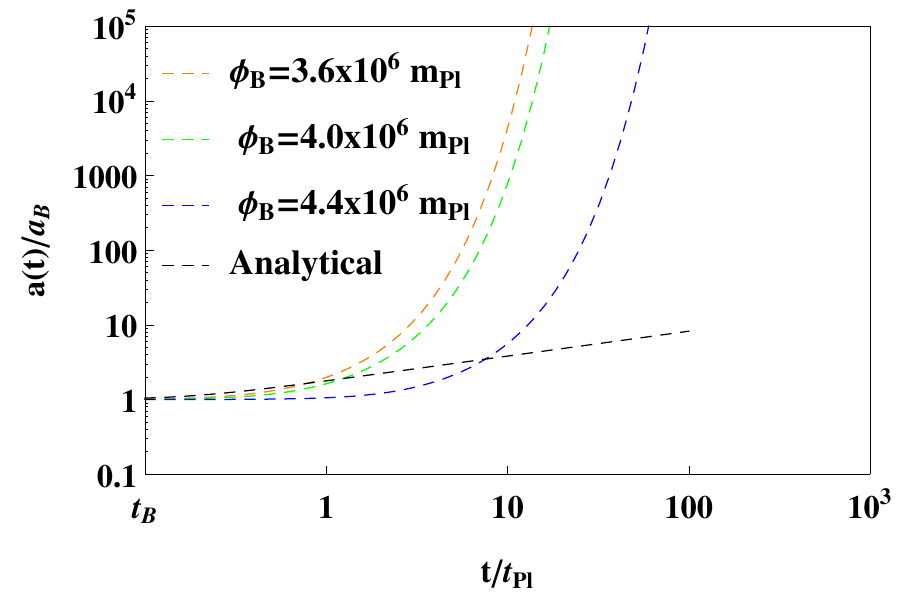}} & 
{\includegraphics[width=2.1in,height=1.6in,angle=0]{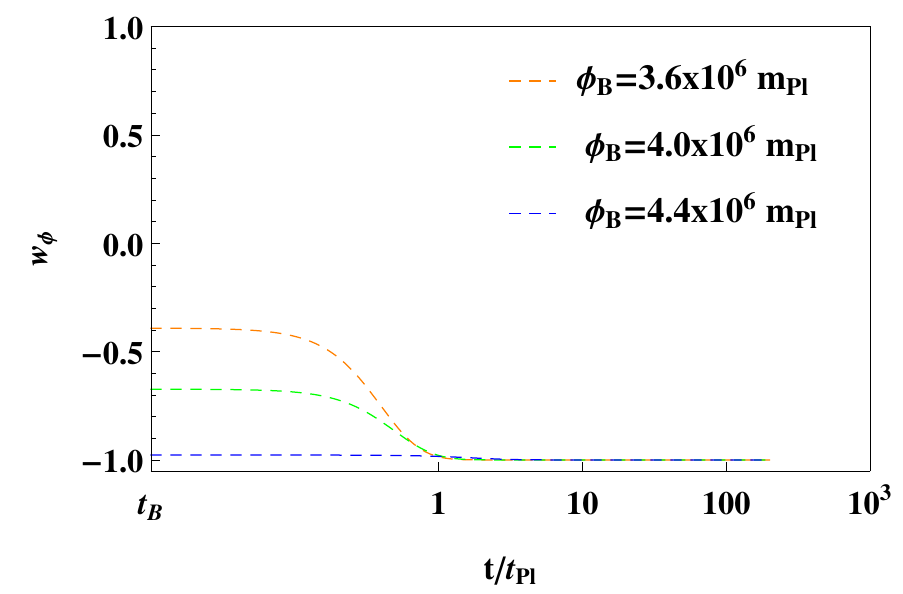}} & 
{\includegraphics[width=2.0in,height=1.6in,angle=0]{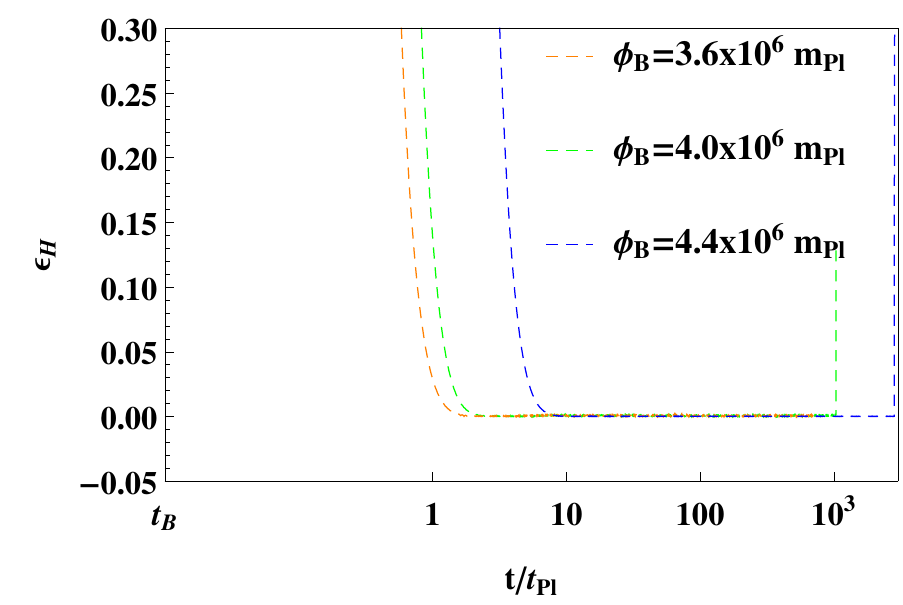}} 
\end{tabular}
\end{center}
\caption{ The numerical solution for the power-law potential with $n=7/4$ and $\dot{\phi}_B>0$. The figure shows the evolution of $a(t)$, $w(\phi)$ and $\epsilon_H$ for the KED (Top panels) and the PED (Bottom panels) initial conditions at the bounce. In the left panels (Top and Bottom), the analytical solution of $a(t)$ [Eq. (\ref{eq:a})] is also exhibited in order to compare it with the numerical results. For the KED case (Top left panel), the evolution of $a(t)$ is universal whereas it is lost in the PED case (Bottom left panel). We use $m=6.2 \times 10^{-6}m_{pl}$ and $m_{pl}=1$.}
\label{fig:7/4a}
\end{figure*}
\begin{figure*}[tbp]
\begin{center}
\begin{tabular}{ccc}
{\includegraphics[width=2.1in,height=1.65in,angle=0]{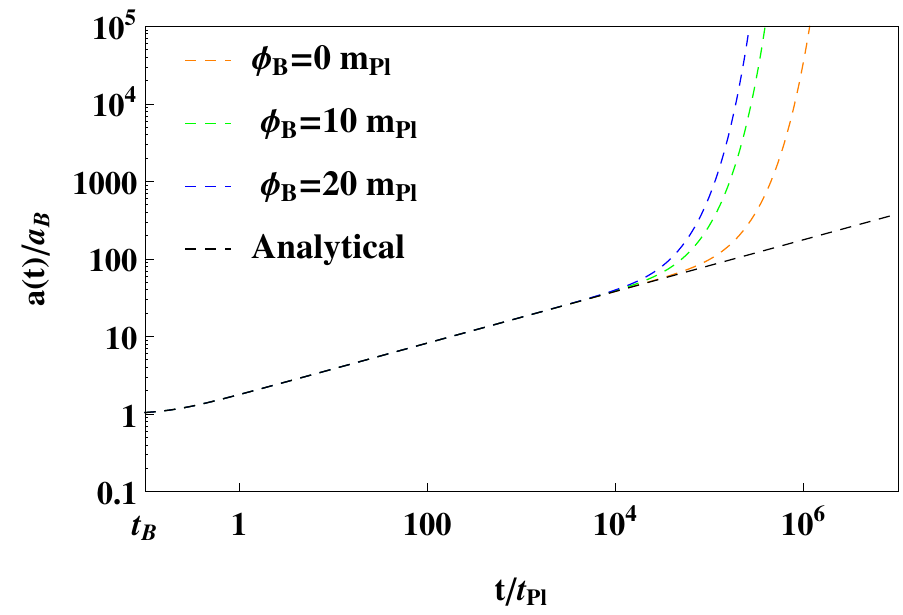}} &
{\includegraphics[width=2.1in,height=1.6in,angle=0]{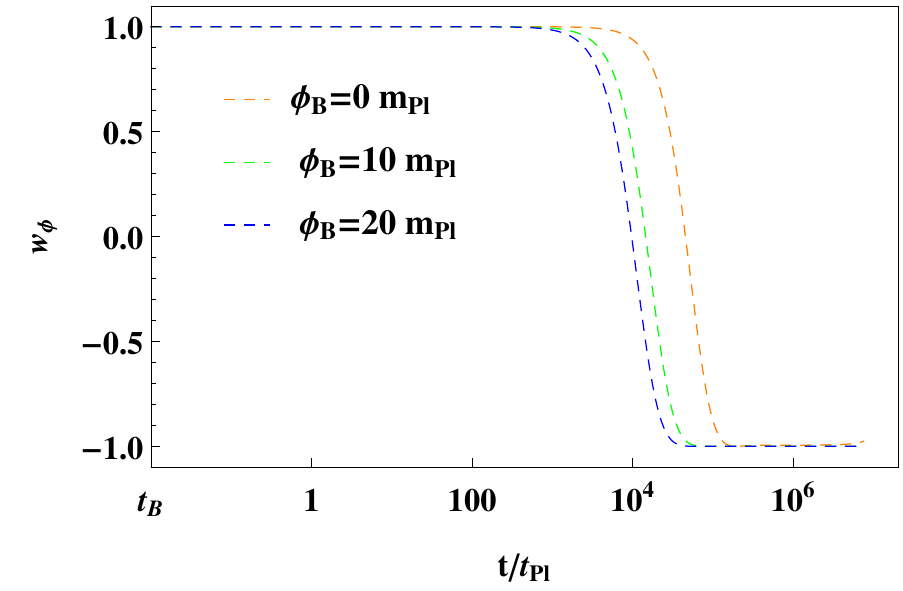}} &
{\includegraphics[width=2.0in,height=1.6in,angle=0]{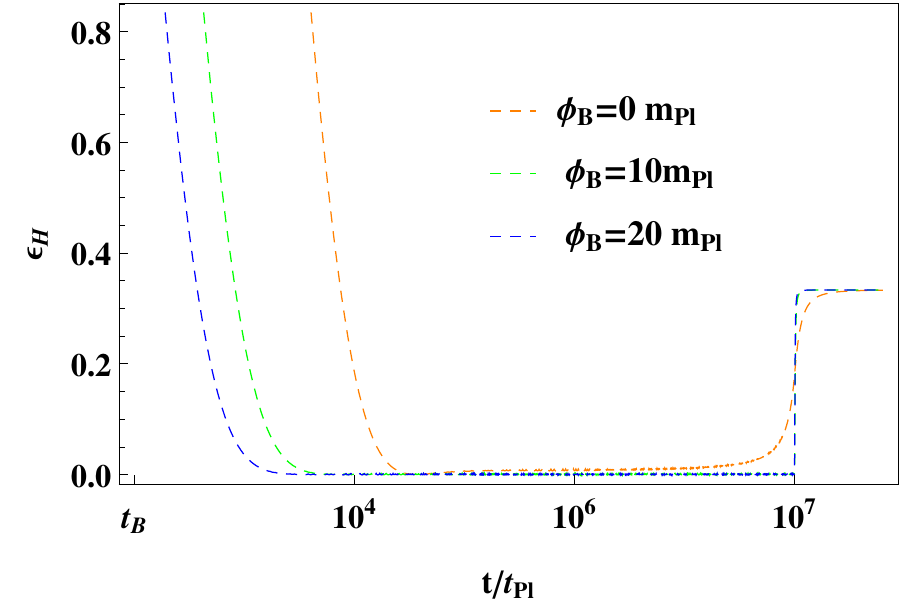}}
 \\
{\includegraphics[width=2.1in,height=1.6in,angle=0]{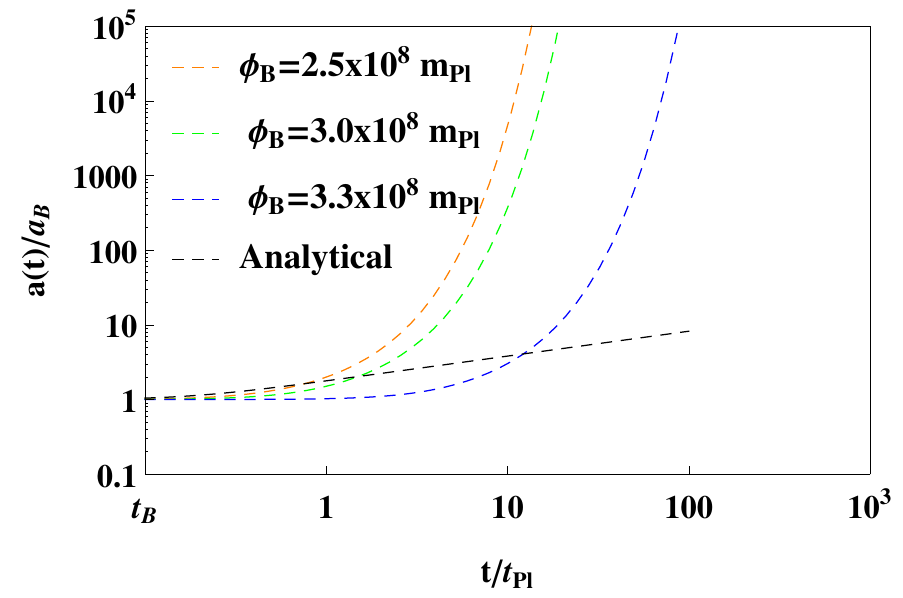}} & 
{\includegraphics[width=2.1in,height=1.6in,angle=0]{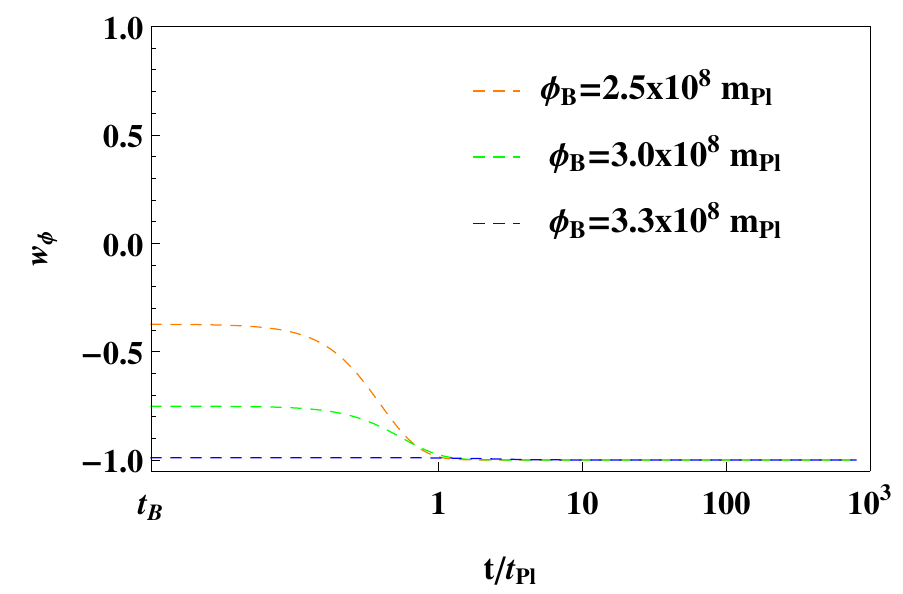}} & 
{\includegraphics[width=2.0in,height=1.6in,angle=0]{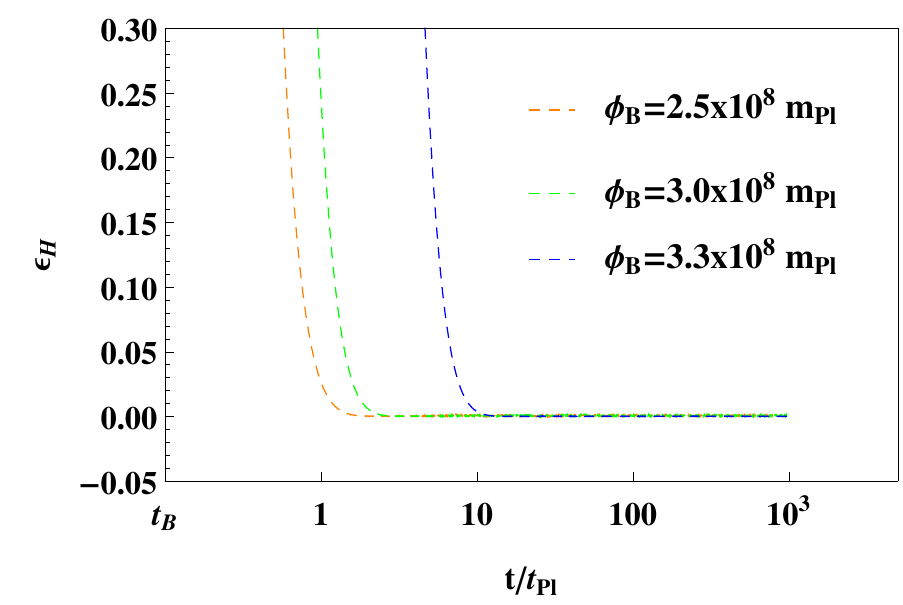}} 
\end{tabular}
\end{center}
\caption{This figure represents the evolution of $a(t)$, $w(\phi)$ and $\epsilon_H$ for the potential with $n=4/3$ and $\dot{\phi}_B>0$. Top panels correspond to KED initial conditions whereas the bottom ones are for PED initial conditions. We take $m=5.1 \times 10^{-5}m_{pl}$ and $m_{pl}=1$.}
\label{fig:4/3a}
\end{figure*}
(3) The number of e-folds $N_{inf}$ during the slow-roll inflation is defined by
\begin{eqnarray}
N_{inf} = ln \Big{(} \frac{a_{end}}{a_i} \Big{)} =  \int_{t_i}^{t_{end}} H(t) dt \nonumber \\
 = \int_{\phi_i}^{\phi_{end}} \frac{H}{\dot{\phi}} d\phi \simeq \int_{\phi_{end}}^{\phi_i} \frac{V}{V_{\phi}} d\phi
\label{eq:Ninf}
\end{eqnarray}
where $a_i$ represents the time when the universe commences to accelerate i.e. $\ddot{a}(t_i) \gtrsim 0$. The $a_{end}$ is the time when the universe ends to accelerate i.e. $w(\phi_{end})=-1/3$.

(4) During the bouncing phase, we can obtain an analytical solution of scale factor $a(t)$ by using equations (\ref{eq:H}) and (\ref{eq:ddphi}). In the bouncing phase, if the effects of potential are negligible compared to the kinetic energy term then equations (\ref{eq:H}) and (\ref{eq:ddphi}) can be written as
\begin{eqnarray}
H^2 &=& \frac{8 \pi}{3 m_{p}^2}~\frac{1}{2}\dot{\phi^2} \Big{(}1-\frac{\dot{\phi^2}}{2\rho_c}\Big{)},\nonumber\\
\ddot{\phi}+3H \dot{\phi}&=&0
\label{eq:Hreduce}
\end{eqnarray}
On solving the above equations analytically, we find
\begin{eqnarray}
\dot{\phi} &=& \pm \sqrt{2 \rho_c} \left( \frac{a_B}{a(t)} \right)^3,\nonumber \\
a(t) &=& a_B \left( 1+ \delta \frac{t^2}{t_{pl}^2} \right)^{1/6}
\label{eq:a}
\end{eqnarray}
where $t_{pl}$  denotes the Planck time, and parameter $\delta = \frac{24 \pi \rho_c}{m_{pl}^4}$ represents a dimensionless constant.

(5) We also introduce a new quantity $r_w$, which is  the ratio between the kinetic and potential energies, 
\begin{eqnarray}
r_{w} &\equiv & \frac{\frac{1}{2}\dot{\phi}^2}{V(\phi)}
\label{eq:rw}
\end{eqnarray}
From which we define $r_{cw}$ as the ratio that corresponds to exactly 60 e-folds during the slow-roll inflation \cite{Planck2015},
 \begin{eqnarray}
r_{cw} &\equiv & \frac{\frac{1}{2}\dot{\phi}^2}{V(\phi)} \Big{\vert}_{N_{inf}\simeq 60}
\label{eq:rwb}
\end{eqnarray}

In the following sub-section, we shall discuss power-law potentials with different $n$ in the context of PIV and NIV at the bounce.
\subsection{Positive inflaton velocity: $\dot{\phi}_B > 0$ }
\label{sec:phiB>0}
For power-law potential (\ref{eq:pot}) with $n<2$, we focus on the positive values of the scalar field $\phi$ as for negative values of $\phi$, potential would be complex (except $n=1$, in this case, potential would be negative). Therefore, $\phi$ must be positive in order to keep the potential to be real.
\subsubsection{Power-law potential with $n=7/4$}
\label{sec:n=7/4a}
Let us start by considering the evolution equations (\ref{eq:H}) and (\ref{eq:ddphi}) with (\ref{eq:pot}) and $n=7/4$. As mentioned above, we choose only positive values of initial conditions of $\phi_B$ in order for the potential to be real. Further, it can be divided into two subclasses, KED and PED cases at the bounce.

Fig. \ref{fig:7/4a} (Top panels) corresponds to the KED initial conditions, for which we evolve the equations (\ref{eq:H}), (\ref{eq:ddphi}) and (\ref{eq:pot}) numerically, and obtain the evolution of the scale factor $a(t)$, the EOS $w(\phi)$ and the slow-roll parameter $\epsilon_H$ for the same set of initial conditions of $\phi_B$. It is clearly exhibited that the desired slow-roll inflation for a set of initial conditions is obtained. During this phase, $a(t)$ is exponentially increasing (Top left panel of Fig. \ref{fig:7/4a}), $w(\phi)$ is nearly close to $-1$ (Top middle panel of Fig. \ref{fig:7/4a}), and  $\epsilon_H \ll 1$ (Top right panel of Fig. \ref{fig:7/4a}).

From the top middle panel of Fig. \ref{fig:7/4a}, we notice that the evolution of the universe can be split into three phases, namely bouncing, transition and the slow-roll \cite{Tao2017}. In the bouncing phase, the kinetic energy is dominated, and $w(\phi) \simeq +1$. In the transition phase ($t/t_{pl}\simeq 10^4$), $w(\phi)$ changes from $+1$ to $-1$ ($t/t_{pl}\simeq 10^5$). In the slow-roll phase, $w(\phi)$ remains to $-1$ till the end of the slow-roll inflation. In the bouncing phase, it is astounding to note that the evolution of $a(t)$ is independent for a wide varieties of initial conditions of $\phi_B$, and can be well approximated by the analytical solution (\ref{eq:a}).

Next, we calculate the number of e-folds $N_{inf}$ during the slow-roll inflation for any choice of $\phi_B$ in the range $(0, \phi_{max})$, and the relevant quantities are shown in Table \ref{tab:7/4a}. For the successful inflation, at least 60 e-folds are needed and to get it, one has to require
\begin{eqnarray}
\phi_B \in (0.805m_{pl}, \phi_{max})
\end{eqnarray}
where
\begin{eqnarray}
\phi_{max} = \left( \frac{16 \rho_c^4}{m^9}  \right)^{1/7} \equiv 4.4284 \times 10^6
\label{eq:phim7/4}
\end{eqnarray}
From Table \ref{tab:7/4a}, we can clearly see that $N_{inf}$ increases as $\phi_B$ increases, this implies that the larger values of $\phi_B$ give rise to more number of e-folds during the slow-roll inflation. The similar results for power-law potential with $n=2$ are shown in \cite{Tao2017}. 

The bottom panels of Fig. \ref{fig:7/4a} are shown for the PED initial conditions at the bounce. In this case, one can clearly notice that the universality of the scale factor $a(t)$ is lost, and the bouncing phase does not exist any more, though the slow-roll inflationary phase $w(\phi) \simeq -1$ can still be obtained. As mentioned above, we get more number of e-folds for the larger values of $\phi_B$, as Table \ref{tab:7/4a} shows.
\begin{figure*}[tbp]
\begin{center}
\begin{tabular}{ccc}
{\includegraphics[width=2.1in,height=1.62in,angle=0]{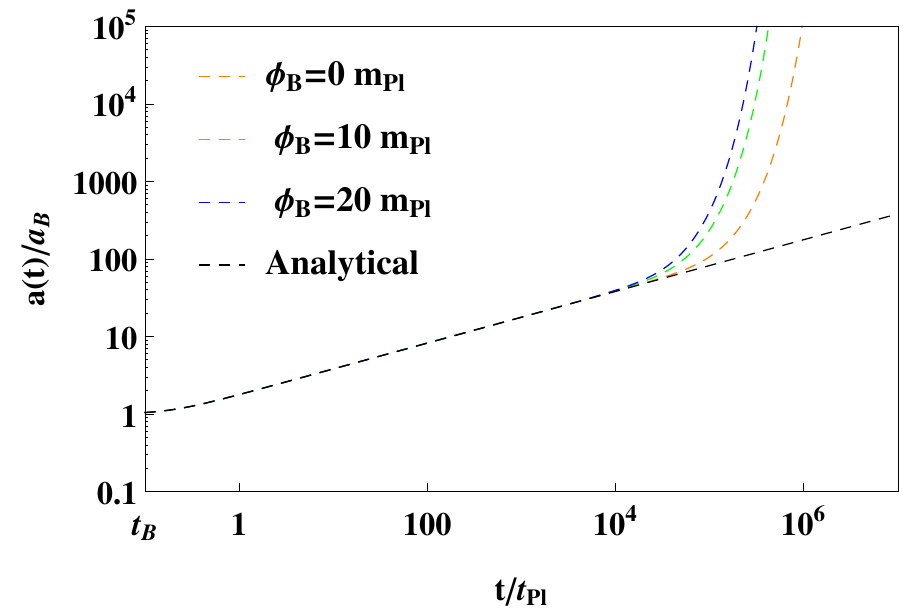}} &
{\includegraphics[width=2.1in,height=1.57in,angle=0]{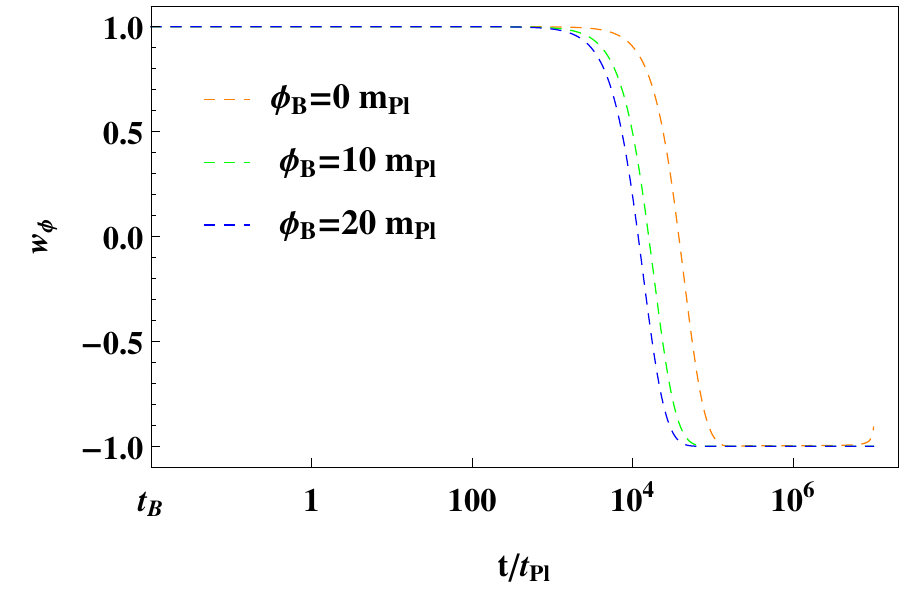}} &
{\includegraphics[width=2.0in,height=1.56in,angle=0]{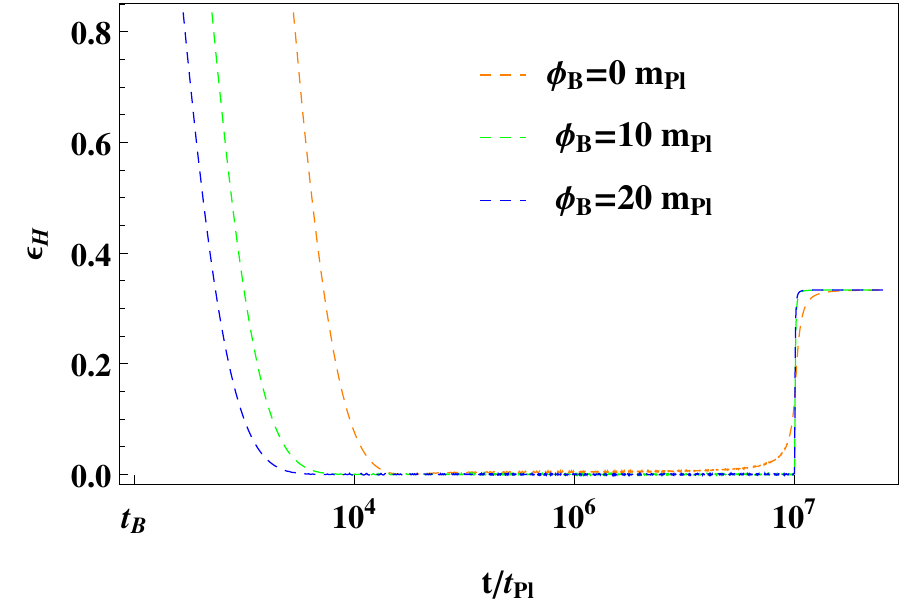}}
 \\
{\includegraphics[width=2.1in,height=1.6in,angle=0]{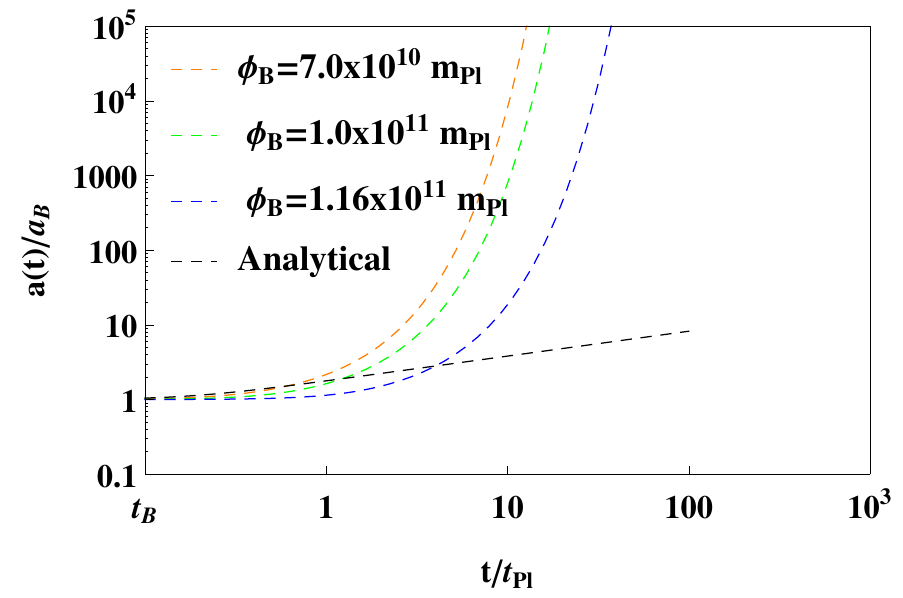}} & 
{\includegraphics[width=2.1in,height=1.6in,angle=0]{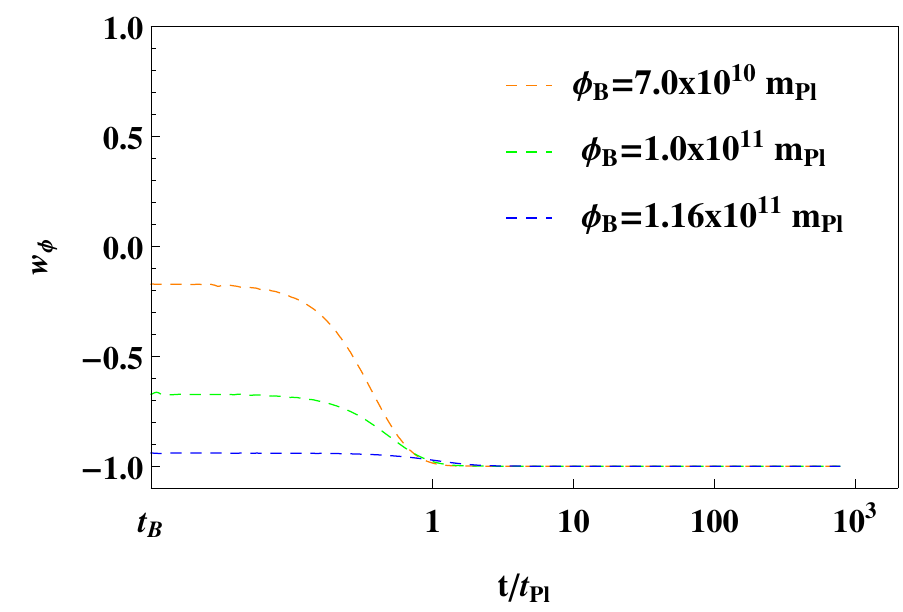}} & 
{\includegraphics[width=2.0in,height=1.6in,angle=0]{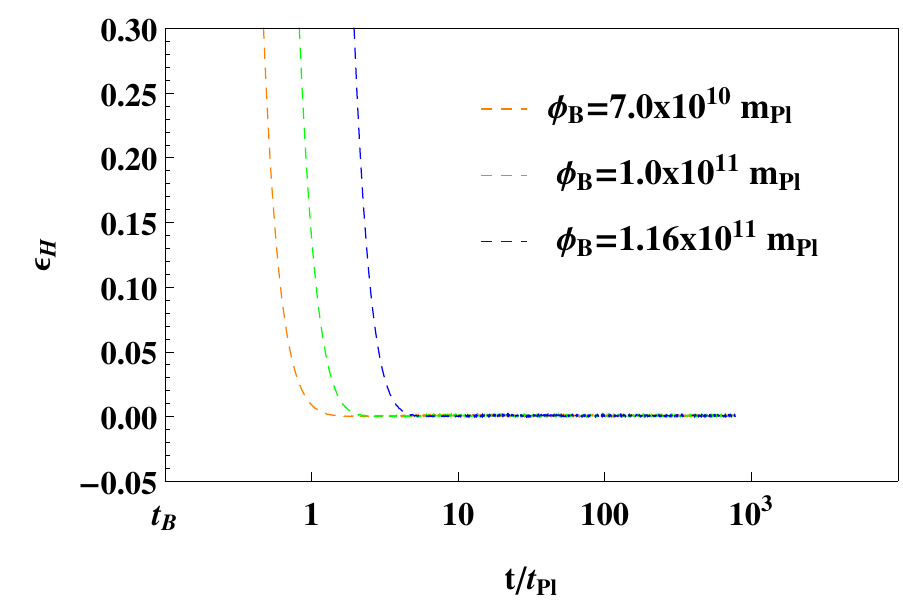}} 
\end{tabular}
\end{center}
\caption{ This figure is displayed for $n=1$ and $\dot{\phi}_B>0$. We choose $m=1.9 \times 10^{-4}m_{pl}$ and $m_{pl}=1$.}
\label{fig:n=1a}
\end{figure*}
\begin{table*}
\caption{Table for the number of e-foldings $N_{inf}$, $r_{cw}(N_{inf} \simeq 60)$ and $r_w(60 < N_{inf} < 60)$ with $\dot{\phi}_B>0$. The values of $\phi_B$ start form the KED case and approach till the end of PED one. In PED initial conditions, we put the sign of $\simeq$ as both values of $\phi_{*}$ are not exactly equal.}
\begin{center}
\resizebox{\textwidth}{!}{%
\begin{tabular}{cccccccc}
\hline\hline
 $n$~~  & $m$ ~~& $\phi_B$~~~  & Inflation~~~ & $t/t_{pl}$~~~ & $\phi_{*}$~~~ & $N_{inf}$ &~~~$r_{cw}/r_w$\\
\hline\hline
\\
7/4 ~~ & $6.2 \times 10^{-6}$~~& 0~~~ & starts~~~& $6.6745 \times 10^4$ ~~~& 2.18488~~~ & 38.5781 &~~~ $r_{cw} > r_w$\\
 ~~&~~&~~~& ends~~~& $1.111 \times 10^7$ ~~~& 0.180914~~~ &  \\\\
~~ &~~& 0.5~~~ & starts~~~& $5.6425 \times 10^4$ ~~~& 2.65752~~~ & 53.7887 &~~~ $r_{cw} > r_w$\\
 ~~&~~&~~~& ends~~~& $1.309 \times 10^7$ ~~~& 0.267712~~~ &  \\\\ 
~~ &~~& 0.795~~~ & starts~~~& $5.17513 \times 10^4$ ~~~& 2.93845~~~ &57.7869 &~~~ $r_{cw} > r_w$\\
 ~~&~~&~~~& ends~~~& $1.0699 \times 10^7$ ~~~& 1.34347~~~ &  \\\\ 
~~ &~~& 0.805~~~ & starts~~~& $5.16106 \times 10^4$ ~~~& 2.948~~~ & 60.3252 &~~~ $r_{cw} = r_w$\\
 ~~&~~&~~~& ends~~~& $1.2024 \times 10^7$ ~~~& 1.5552~~~ &  \\\\  
~~ &~~& 1.0~~~ & starts~~~& $4.8950 \times 10^4$ ~~~& 3.13439~~~ & 63.7207 &~~~ $r_{cw} < r_w$\\
~~&~~&~~~& ends~~~& $1.119 \times 10^7$ ~~~& 1.67068~~~ &  \\\\   
~~&~~&3.0~~~& starts~~~& $3.23156 \times 10^4$ ~~~& 5.06678~~~ & 105.443 &~~~ $r_{cw} < r_w$\\
   ~~&~~&~~~& ends~~~& $1.0402 \times 10^7$ ~~~& 3.74002~~~ &  \\\\   
~~ &~~& (PED)$4 \times 10^6$~~~ & starts~~~& 0.577 ~~~& $\simeq4 \times 10^6$~~~ & 714.0459 &~~~ $r_{cw} < r_w$\\
~~&~~&~~~& ends~~~& 1047.32 ~~~& $\simeq4 \times 10^6$~~~ &  \\ \\  
4/3 ~~ & $5.1 \times 10^{-5}$~~& 0~~~ & starts~~~& $5.81699 \times 10^4$ ~~~& 2.16248~~~ & 50.4897 &~~~ $r_{cw} > r_w$\\
 ~~&~~&~~~& ends~~~& $1.16825 \times 10^7$ ~~~& 0.11362~~~ &  \\\\
~~ & ~~& 0.35~~~ & starts~~~& $5.29494 \times 10^4$ ~~~& 2.49717~~~ & 57.8107 &~~~ $r_{cw} > r_w$\\
 ~~&~~&~~~& ends~~~& $1.02966 \times 10^7$ ~~~& 0.937812~~~ &  \\\\  
~~ & ~~& 0.4005~~~ & starts~~~& $5.22858 \times 10^4$ ~~~& 2.54562~~~ & 60.0079 &~~~ $r_{cw} = r_w$\\
 ~~&~~&~~~& ends~~~& $1.0607 \times 10^7$ ~~~& 1.04617~~~ &  \\\\ 
~~ & ~~& 0.5~~~ & starts~~~& $5.10331 \times 10^4$ ~~~& 2.64118~~~ & 62.4811 &~~~ $r_{cw} < r_w$\\
 ~~&~~&~~~& ends~~~& $1.06214 \times 10^7$ ~~~& 0.856514~~~ &  \\\\   
~~ & ~~& 1.0~~~ & starts~~~& $4.57248 \times 10^4$ ~~~& 3.12329~~~ & 71.8501 &~~~ $r_{cw} < r_w$\\
 ~~&~~&~~~& ends~~~& $1.01714 \times 10^7$ ~~~& 1.76734~~~ &  \\\\ 
~~ & ~~& 3.0~~~ & starts~~~& $3.32295 \times 10^4$ ~~~& 5.07132~~~ & 107.173 &~~~ $r_{cw} < r_w$\\
 ~~&~~&~~~& ends~~~& $1.00928 \times 10^7$ ~~~& 4.00218~~~ &  \\\\  
~~ &~~& (PED)$2.5 \times 10^8$~~~ & starts~~~& 0.401 ~~~& $\simeq2.5 \times 10^8$~~~ & 710.9396 &~~~ $r_{cw} < r_w$\\
~~&~~&~~~& ends~~~& 827.735 ~~~& $\simeq2.5 \times 10^8$~~~ &  \\ \\   
~~ &~~&  (PED)$3.0 \times 10^8$~~~ & starts~~~& 0.665 ~~~& $\simeq3.0 \times 10^8$~~~ & 714.4861 &~~~ $r_{cw} < r_w$\\
~~&~~&~~~& ends~~~& $1.17643 \times 10^3$~~~& $\simeq3.0 \times 10^8$~~~ &  \\   
\hline\hline
\end{tabular}}
\label{tab:7/4a}
\end{center}
\end{table*}
\begin{figure*}[tbp]
\begin{center}
\begin{tabular}{ccc}
{\includegraphics[width=2.1in,height=1.6in,angle=0]{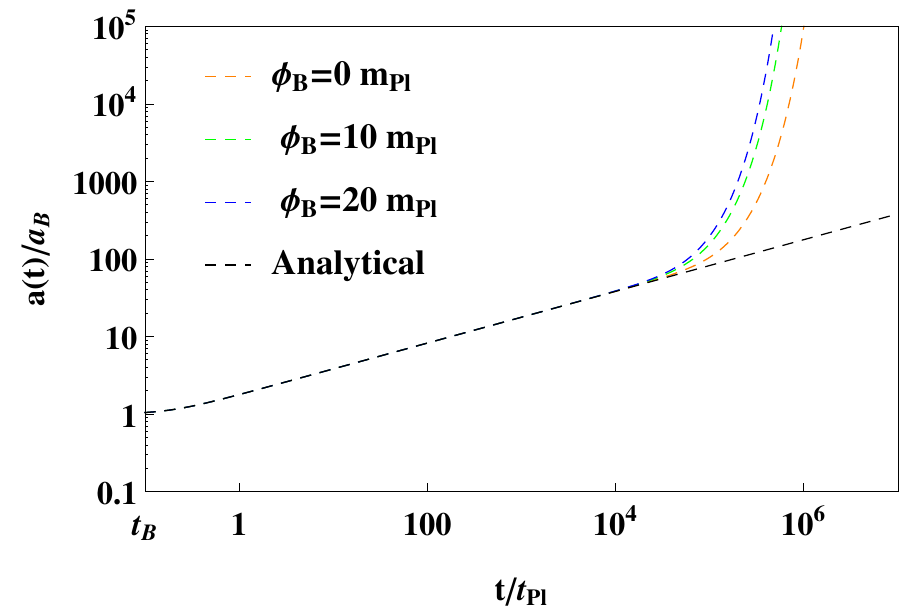}} &
{\includegraphics[width=2.1in,height=1.56in,angle=0]{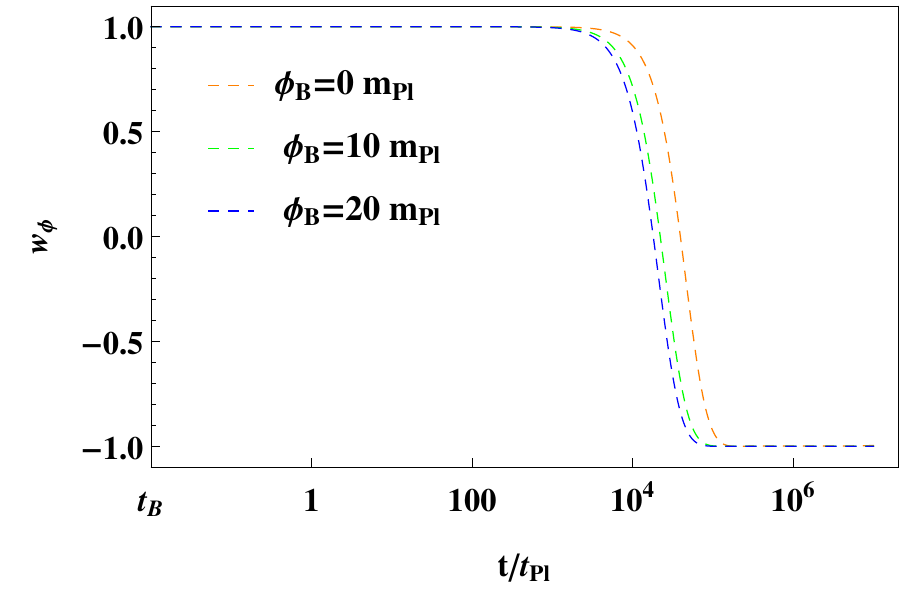}} &
{\includegraphics[width=2.0in,height=1.6in,angle=0]{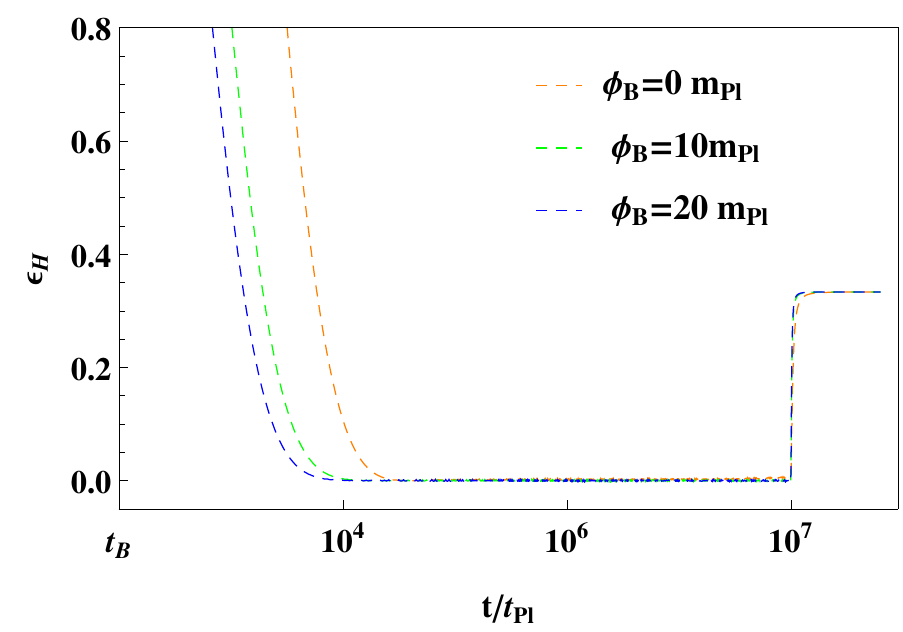}}
 \\
{\includegraphics[width=2.1in,height=1.6in,angle=0]{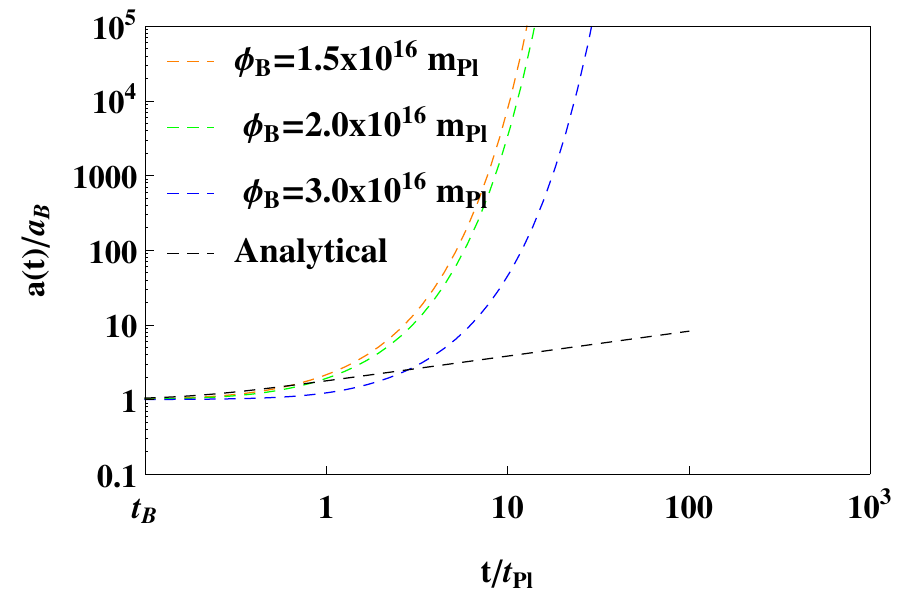}} & 
{\includegraphics[width=2.1in,height=1.61in,angle=0]{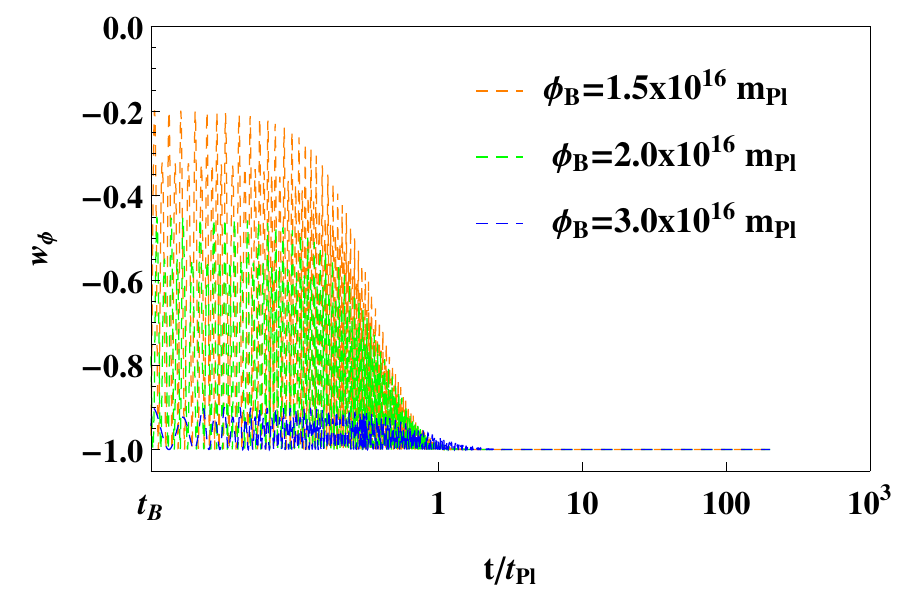}} & 
{\includegraphics[width=2.0in,height=1.6in,angle=0]{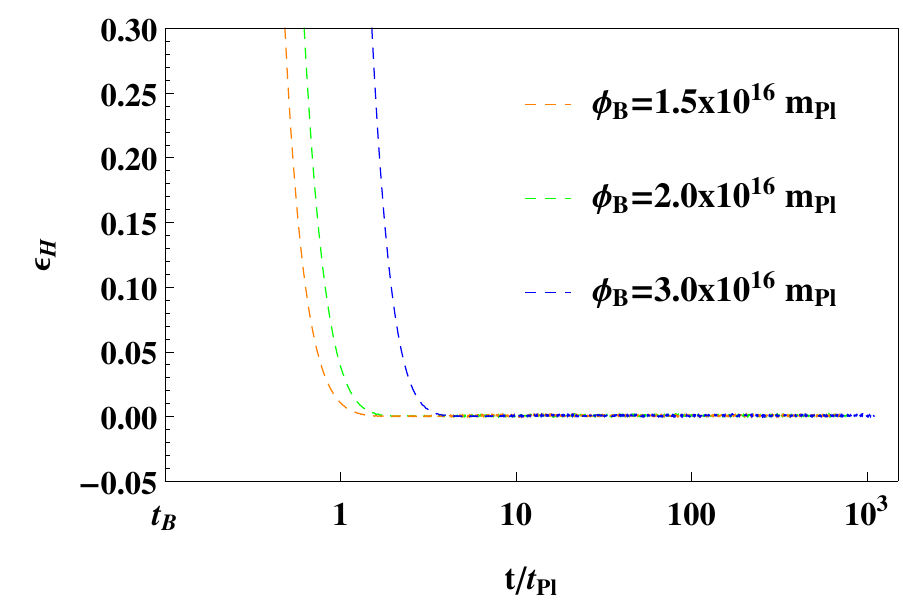}} 
\end{tabular}
\end{center}
\caption{ This figure exhibits the numerical evolution of $a(t)$, $w(\phi)$ and $\epsilon_H$ with $n=2/3$ and $\dot{\phi}_B>0$ for the different sets of initial conditions. We consider $m=4.7 \times 10^{-4}m_{pl}$ and $m_{pl}=1$.}
\label{fig:2/3a}
\end{figure*}
\begin{figure*}[tbp]
\begin{center}
\begin{tabular}{ccc}
{\includegraphics[width=2.1in,height=1.6in,angle=0]{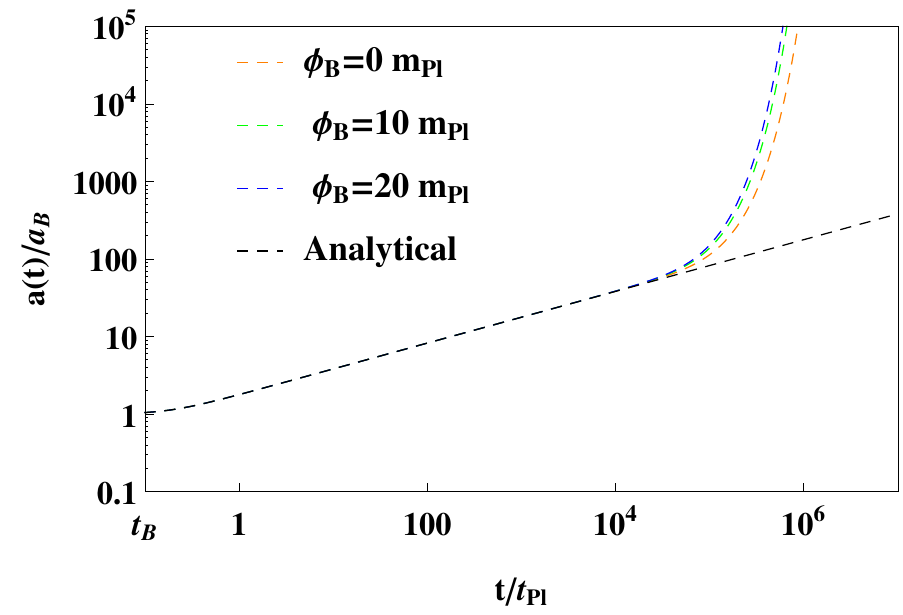}} &
{\includegraphics[width=2.1in,height=1.55in,angle=0]{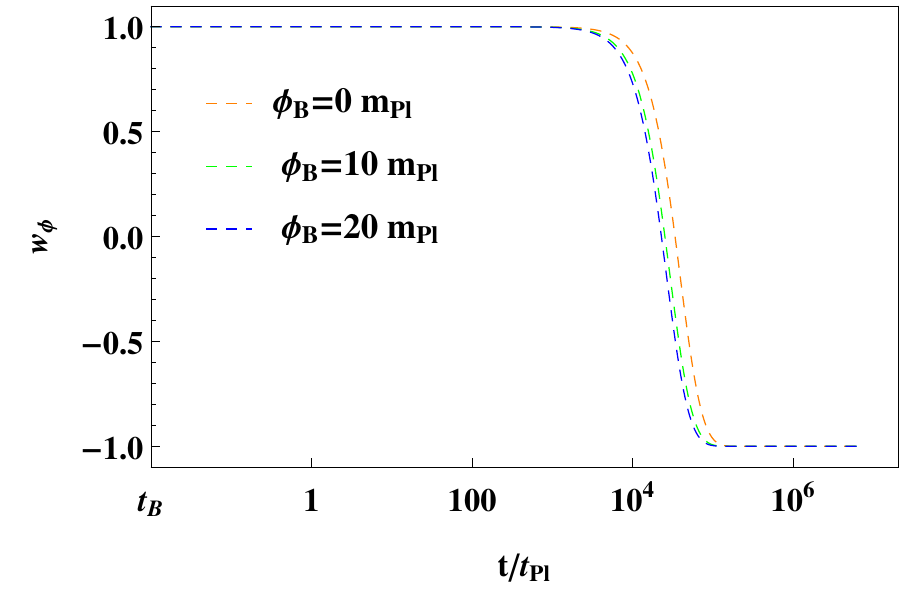}} &
{\includegraphics[width=2.0in,height=1.6in,angle=0]{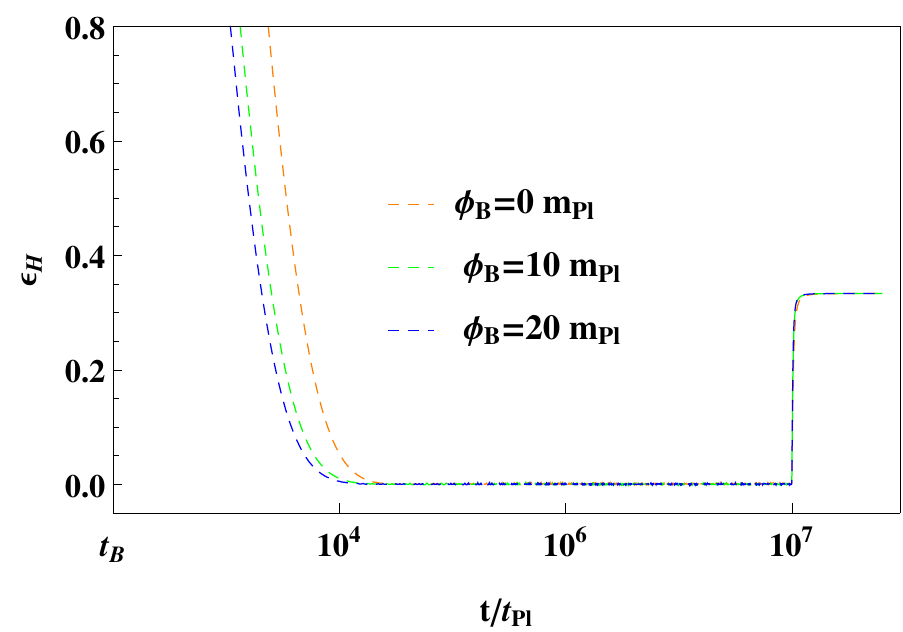}}
 \\
{\includegraphics[width=2.1in,height=1.6in,angle=0]{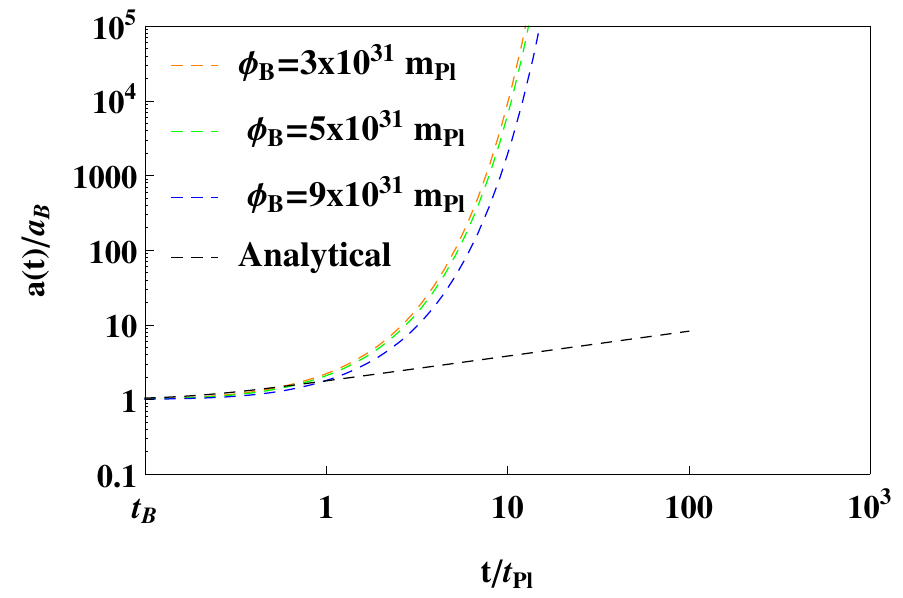}} & 
{\includegraphics[width=2.1in,height=1.61in,angle=0]{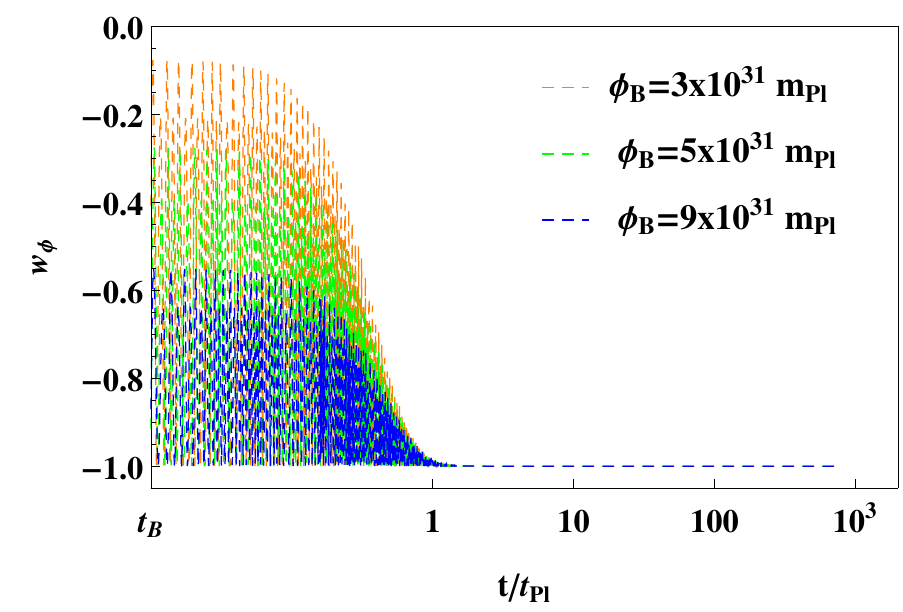}} & 
{\includegraphics[width=2.0in,height=1.57in,angle=0]{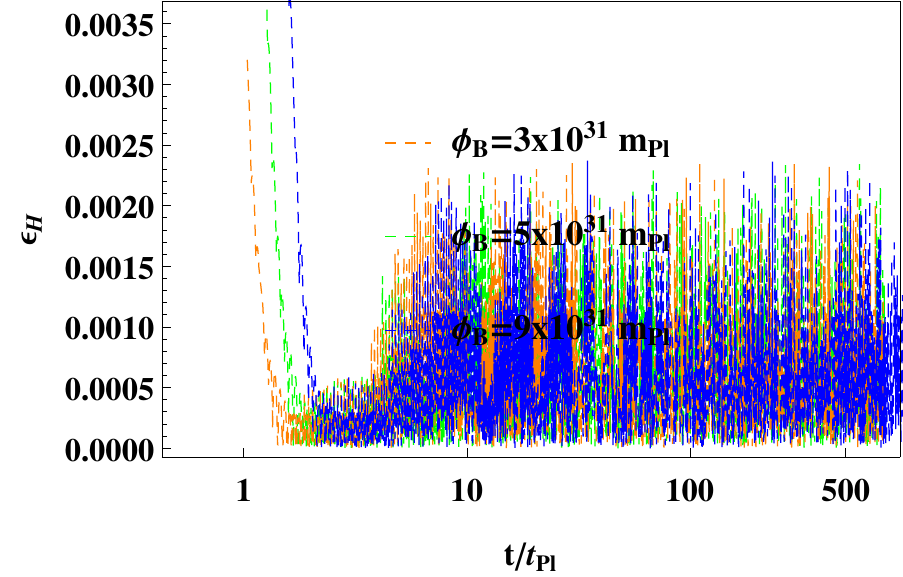}} 
\end{tabular}
\end{center}
\caption{ This figure is plotted for $n=1/3$ and $\dot{\phi}_B>0$. Here, $m=1.1 \times 10^{-3}m_{pl}$ and $m_{pl}=1$.}
\label{fig:1/3a}
\end{figure*}
\begin{table*}
\caption{Table for $N_{inf}$, $r_{cw}(N_{inf} \simeq 60)$ and $r_w(60 < N_{inf} < 60)$ with $\dot{\phi}_B>0$.}
\begin{center}
\resizebox{\textwidth}{!}{%
\begin{tabular}{cccccccc}
\hline\hline
 $n$~~  & $m$ ~~& $\phi_B$~~~  & Inflation~~~ & $t/t_{pl}$~~~ & $\phi_{*}$~~~ & $N_{inf}$ &~~~$r_{cw}/r_w$\\
\hline\hline
\\
1 ~~ & $1.9 \times 10^{-4}$~~& 0~~~ & starts~~~& $4.82718 \times 10^4$ ~~~& 2.13211~~~ & 63.6793 &~~~ $r_{cw} < r_w$\\
 ~~&~~&~~~& ends~~~& $1.04513 \times 10^7$ ~~~& 0.152698~~~ &  \\\\
~~ & ~~& 0.01~~~ & starts~~~& $4.8161 \times 10^4$ ~~~& 2.14174~~~ & 64.7404 &~~~ $r_{cw} < r_w$\\
 ~~&~~&~~~& ends~~~& $1.07438 \times 10^7$ ~~~& 0.075339~~~ &  \\\\ 
~~ & ~~& 0.05~~~ & starts~~~& $4.774 \times 10^4$ ~~~& 2.18031~~~ & 66.0862 &~~~ $r_{cw} < r_w$\\
 ~~&~~&~~~& ends~~~& $1.06922 \times 10^7$ ~~~& 0.169875~~~ &  \\\\  
~~ & ~~& 0.1~~~ & starts~~~& $4.72339 \times 10^4$ ~~~& 2.22857~~~ & 69.9625 &~~~ $r_{cw} < r_w$\\
 ~~&~~&~~~& ends~~~& $1.18715 \times 10^7$ ~~~& 0.574322~~~ &  \\\\   
~~ & ~~& 0.5~~~ & starts~~~& $4.36669 \times 10^4$ ~~~& 2.61579~~~ & 78.0678 &~~~ $r_{cw} < r_w$\\
 ~~&~~&~~~& ends~~~& $1.0538 \times 10^7$ ~~~& 0.929275~~~ &  \\\\ 
~~ &~~& (PED)$1.0 \times 10^{11}$~~~ & starts~~~& 0.575 ~~~& $\simeq1.0 \times 10^{11}$~~~ & 714.3224 &~~~ $r_{cw} < r_w$\\
~~&~~&~~~& ends~~~& $1.0474 \times 10^3$ ~~~& $\simeq1.0 \times 10^{11}$~~~ &  \\ \\  
2/3 ~~ & $4.7 \times 10^{-4}$~~& 0~~~ & starts~~~& $5.05957 \times 10^4$ ~~~& 2.13979~~~ & 72.75 &~~~ $r_{cw} < r_w$\\
 ~~&~~&~~~& ends~~~& $1.06725 \times 10^7$ ~~~& 0.843284~~~ &  \\\\
~~ & ~~& 0.01~~~ & starts~~~& $5.05283 \times 10^4$ ~~~& 2.14957~~~ & 73.1079 &~~~ $r_{cw} < r_w$\\
 ~~&~~&~~~& ends~~~& $1.07272 \times 10^7$ ~~~& 1.44494~~~ &  \\\\ 
~~ & ~~& 0.1~~~ & starts~~~& $4.9876 \times 10^4$ ~~~& 2.23745~~~ & 74.2893 &~~~ $r_{cw} < r_w$\\
 ~~&~~&~~~& ends~~~& $1.0638 \times 10^7$ ~~~& 0.998999~~~ &  \\\\ 
~~ & ~~& 1.0~~~ & starts~~~& $4.47227 \times 10^4$ ~~~& 3.11971~~~ & 85.8929 &~~~ $r_{cw} < r_w$\\
 ~~&~~&~~~& ends~~~& $1.06634 \times 10^7$ ~~~& 2.79913~~~ &  \\\\ 
~~ &~~& (PED)$1.5 \times 10^{16}$~~~ & starts~~~& 0.343 ~~~& $\simeq1.5 \times 10^{16}$~~~ & 715.7368 &~~~ $r_{cw} < r_w$\\
~~&~~&~~~& ends~~~& 796.18 ~~~& $\simeq1.5 \times 10^{16}$~~~ &  \\\\   
~~ &~~& (PED)$3.0 \times 10^{16}$~~~ & starts~~~& 1.06 ~~~& $\simeq3.0 \times 10^{16}$~~~ & 716.2100 &~~~ $r_{cw} < r_w$\\
~~&~~&~~~& ends~~~& 1805.14 ~~~& $\simeq3.0 \times 10^{16}$~~~ &  \\\\   
1/3 ~~ & $1.1 \times 10^{-3}$~~& 0~~~ & starts~~~& $4.35782 \times 10^4$ ~~~& 2.11548~~~ & 86.5305 &~~~ $r_{cw} < r_w$\\
 ~~&~~&~~~& ends~~~& $1.00983 \times 10^7$ ~~~& 1.57308~~~ &  \\\\
~~ & ~~& 0.01~~~ & starts~~~& $4.36 \times 10^4$ ~~~& 2.12553~~~ & 88.5284 &~~~ $r_{cw} < r_w$\\
 ~~&~~&~~~& ends~~~& $1.03913 \times 10^7$ ~~~& 1.40504~~~ &  \\\\ 
~~ & ~~& 2.0~~~ & starts~~~& $3.92 \times 10^4$ ~~~& 4.09819~~~ & 97.6823 &~~~ $r_{cw} < r_w$\\
 ~~&~~&~~~& ends~~~& $1.00755 \times 10^7$ ~~~& 3.8348~~~ &  \\\\   
~~ &~~&(PED)$3.0 \times 10^{31}$~~~ & starts~~~& 0.3109 ~~~& $\simeq3.0 \times 10^{31}$~~~ & 715.3383 &~~~ $r_{cw} < r_w$\\
~~&~~&~~~& ends~~~& 781.16 ~~~& $\simeq3.0 \times 10^{31}$~~~ &  \\  
\hline\hline
\end{tabular}}
\label{tab:2/3a}
\end{center}
\end{table*}
\subsubsection{Power-law potential with $n=4/3$}
\label{sec:n=4/3a}
In this sub-section, we consider power-law potential with $n=4/3$. Similar to the case of $n=7/4$, we choose only positive values of the scalar field to get the real potential. Later, the initial conditions at the quantum bounce can be divided into two categories such as KED and PED initial conditions.

Let us first focus on the case in which the evolution at the quantum bounce is dominated by the kinetic energy of the inflaton field. To this effect, we numerically evolve the system (\ref{eq:H}) and (\ref{eq:ddphi}) with (\ref{eq:pot}) and $n=4/3$. The results are shown in the top panels of Fig. \ref{fig:4/3a}. The evolution of analytical solution of
$a(t)$ (\ref{eq:a}) is also illustrated to compare it with the numerical solutions, and found to be universal (Top left panel of Fig. \ref{fig:4/3a}).

As we discussed in the case of $n=7/4$, here also the evolution of the universe is divided into three phases; bouncing, transition, and the slow-roll (Top middle panel of Fig. \ref{fig:4/3a}). In the bouncing phase, the evolution of $a(t)$ is independent not only on the different kind of initial values of $\phi_B$ but also the inflationary potential. This is mainly due to the small amplitude of the potential in comparing with the kinetic term, and its effects on the evolution during the bouncing phase is almost insignificant.

Top right panel of Fig. \ref{fig:4/3a} demonstrates the evolution of the slow-roll parameter $\epsilon_H$. The slow-roll inflation regime can be obtained for any choices of $\phi_B$ in the range $(0, \phi_{max})$ (see Table \ref{tab:7/4a}). However, in order to get at least 60 e-folds during the slow-roll regime, the values of $\phi_B$ should be in the following range
\begin{eqnarray}
\phi_B \in (0.4005m_{pl}, \phi_{max})
\end{eqnarray}
where
\begin{eqnarray}
\phi_{max} = \left( \frac{8 \rho_c^3}{m^8}  \right)^{1/4} \equiv 3.31299 \times 10^8
\label{eq:phim4/3}
\end{eqnarray}
Bottom panels of Fig. \ref{fig:4/3a} represent the evolution of $a(t)$, $w(\phi)$ and $\epsilon_H$ for the PED initial conditions at the bounce. In this case, the bouncing phase no longer exists and the universality of the scale factor disappears. However, slow-roll inflationary phase can still be achieved. Similar to the case of $n=7/4$, as $\phi_B$ increases we get more and more number of e-folds. In other words, the PED initial conditions can produce a large number of e-folds $N_{inf}$ during the slow-roll inflationary phase (see Table \ref{tab:7/4a}).
\subsubsection{Power-law potential with $n=1, 2/3$ and $1/3$}
\label{sec:n=1a}
In this sub-section, we shall study the cases $n=1, 2/3$ and $1/3$ for the power-law potential. Similar to the sub-sections \ref{sec:n=7/4a} and \ref{sec:n=4/3a}, for all the three cases, $\phi_B$ must be positive to obtain the real potential. Further the initial conditions at the quantum bounce can be divided into two categories, namely KED and PED.

Let us first consider the KED case at the bounce. We numerically solve the equations (\ref{eq:H}) and (\ref{eq:ddphi}) with (\ref{eq:pot}) and $n=1, 2/3$ and $1/3$, respectively. The results are displayed in the top panels of Figs. \ref{fig:n=1a}, \ref{fig:2/3a} and \ref{fig:1/3a}, respectively. In the top left panels, the evolution of $a(t)$ at the bounce are independent of the different sets of initial conditions and represent a universal feature. We also depict the analytical solution of $a(t)$ (\ref{eq:a}) in comparison to the numerical solutions. As mentioned in sub-sections \ref{sec:n=7/4a} and \ref{sec:n=4/3a}, here also the evolution of the universe for $n=1, 2/3$ and $1/3$ can be split into three regimes: bouncing, transition and the slow-roll. The corresponding number of e-folds $N_{inf}$ for each case are shown in Table \ref{tab:2/3a}.

For these cases, the desired slow-roll inflation is achieved for any values of $\phi_B$ in the range $(0, \phi_{max})$. In the said range, we also have more than 60 e-folds in contrast to the cases of $n=7/4$ and $4/3$ where to obtain at least 60 e-folds the ranges were restricted.

For the different cases of $n$, the $\phi_{max}$ are found to be:\\
\\
For $n=1$,
\begin{eqnarray}
\phi_{max} = \frac{2 \rho_c}{m^3} \simeq 1.19551 \times 10^{11}
\label{eq:phimn1}
\end{eqnarray} 
for $n=2/3$,
\begin{eqnarray}
\phi_{max} = \left(\frac{8 \rho_c^3}{m^{10}}\right)^{1/2} \simeq 3.23766 \times 10^{16}
\label{eq:phim2/3}
\end{eqnarray}
and for $n=1/3$,
\begin{eqnarray}
\phi_{max} = \frac{8 \rho_c^3}{m^{11}} \simeq 1.1932511 \times 10^{32}
\label{eq:phim1/3}
\end{eqnarray}
\begin{figure*}[tbp]
\begin{center}
\begin{tabular}{ccc}
{\includegraphics[width=2.1in,height=1.6in,angle=0]{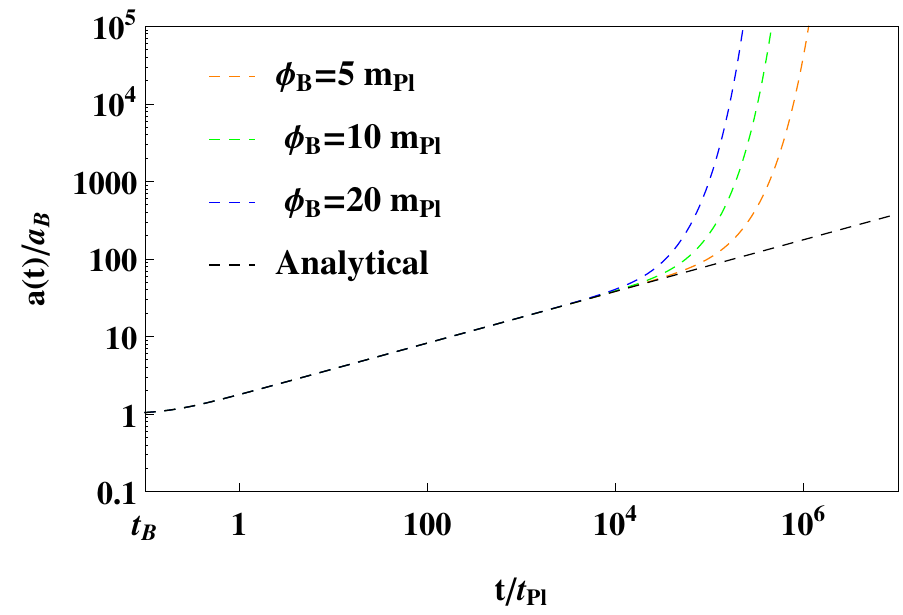}} &
{\includegraphics[width=2.1in,height=1.55in,angle=0]{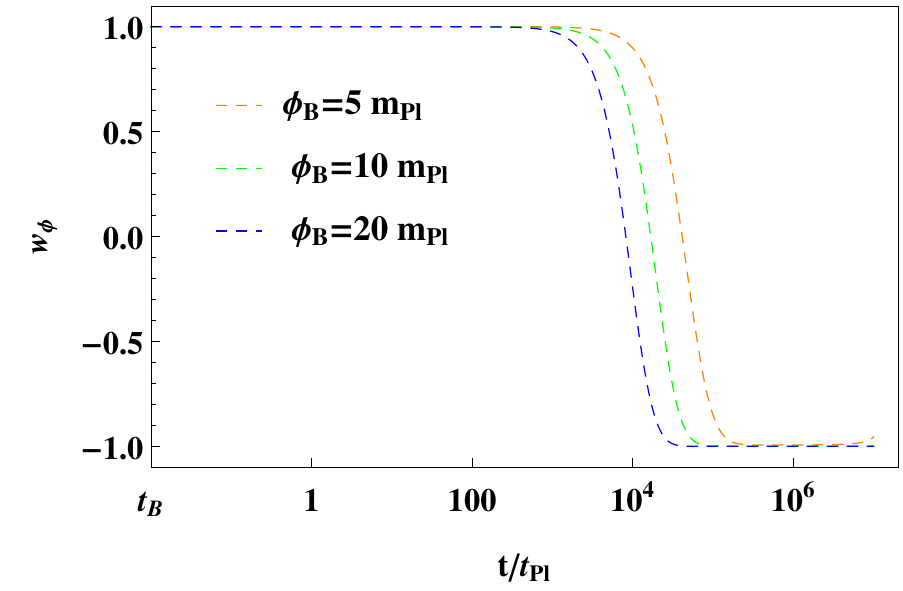}} &
{\includegraphics[width=2.0in,height=1.6in,angle=0]{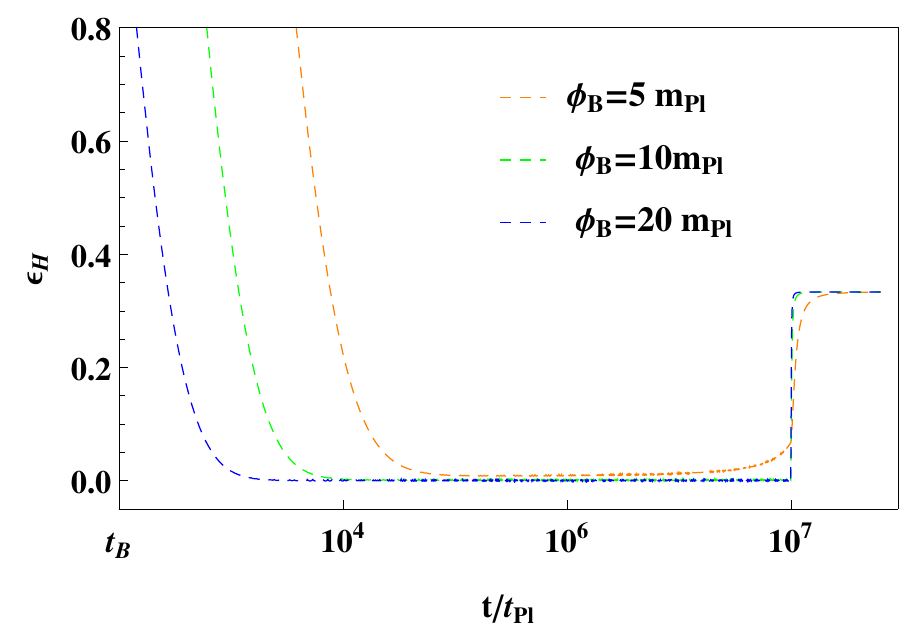}}
 \\
{\includegraphics[width=2.1in,height=1.6in,angle=0]{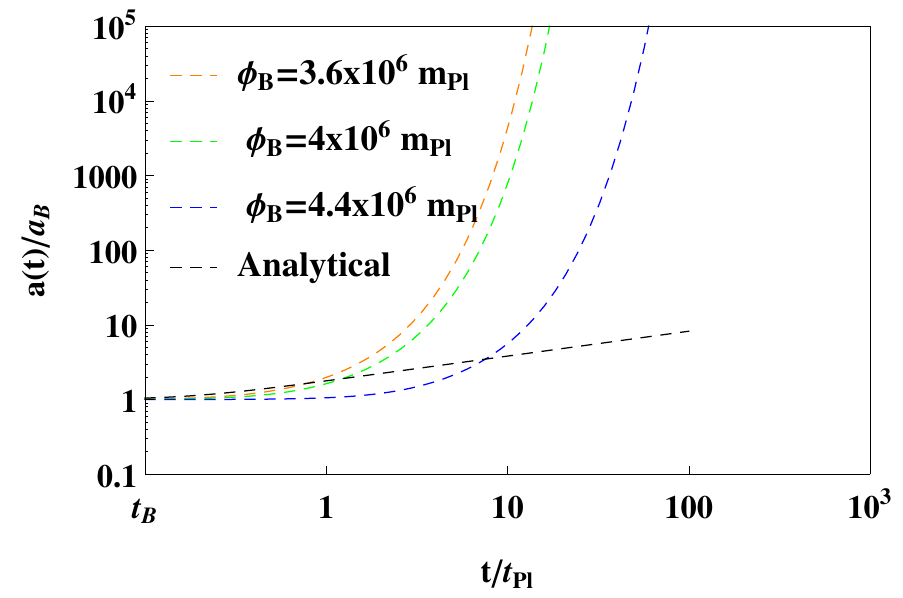}} & 
{\includegraphics[width=2.1in,height=1.6in,angle=0]{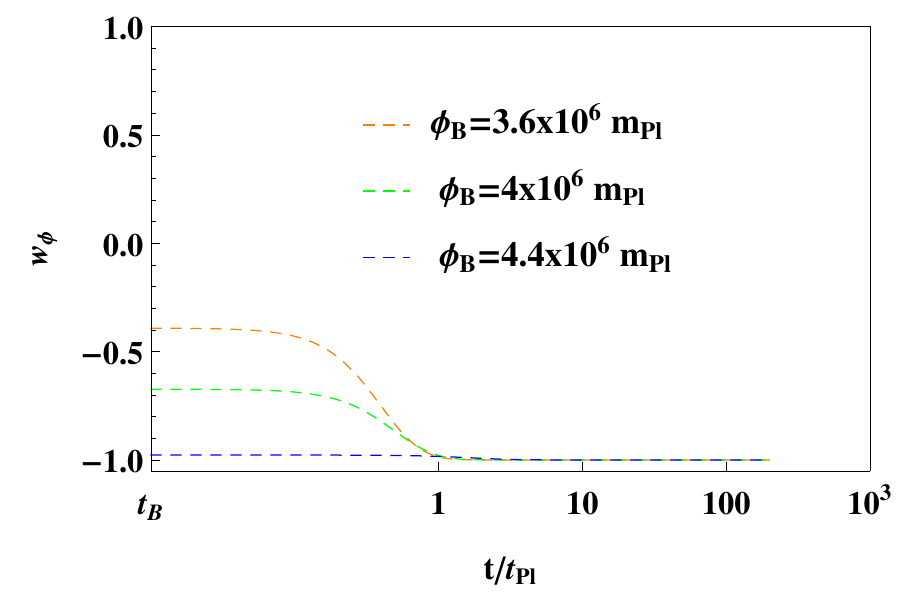}} & 
{\includegraphics[width=2.0in,height=1.6in,angle=0]{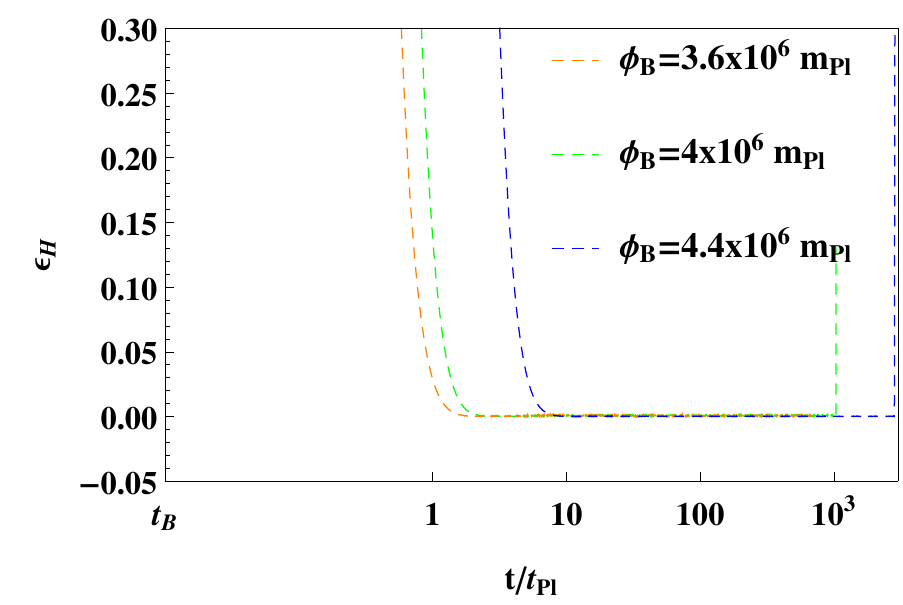}} 
\end{tabular}
\end{center}
\caption{ This figure shows the numerical evolution of $a(t)$, $w(\phi)$ and $\epsilon_H$ for the potential with $n=7/4$ and $\dot{\phi}_B<0$. Top panels demonstrate the evolution for KED initial conditions whereas bottom ones are for PED.}
\label{fig:7/4b}
\end{figure*}
\begin{figure*}[tbp]
\begin{center}
\begin{tabular}{ccc}
{\includegraphics[width=2.1in,height=1.6in,angle=0]{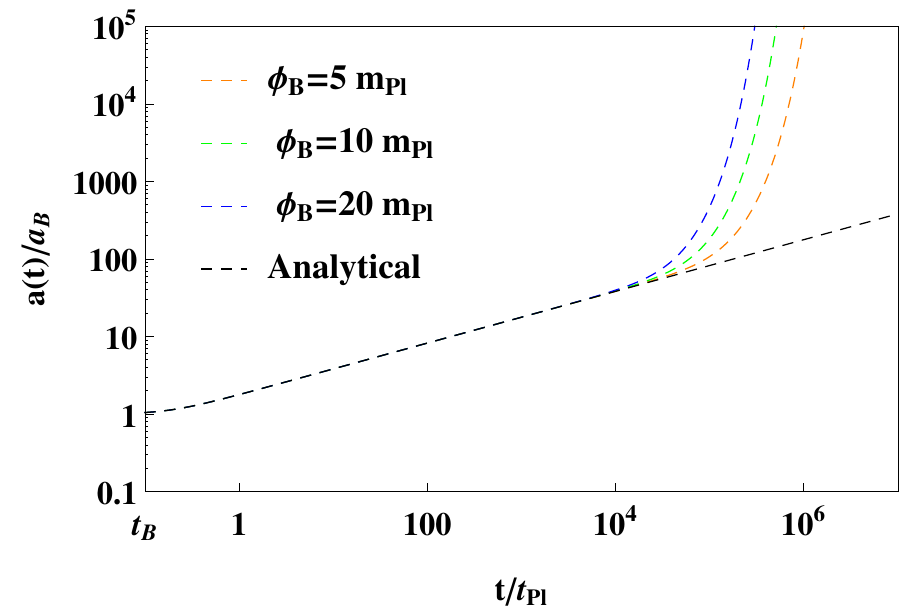}} &
{\includegraphics[width=2.1in,height=1.55in,angle=0]{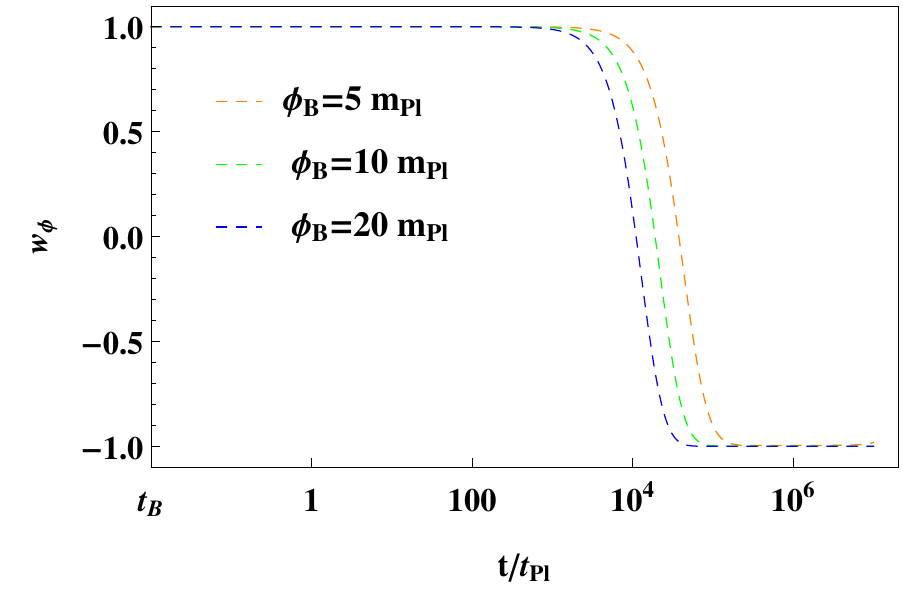}} &
{\includegraphics[width=2.0in,height=1.6in,angle=0]{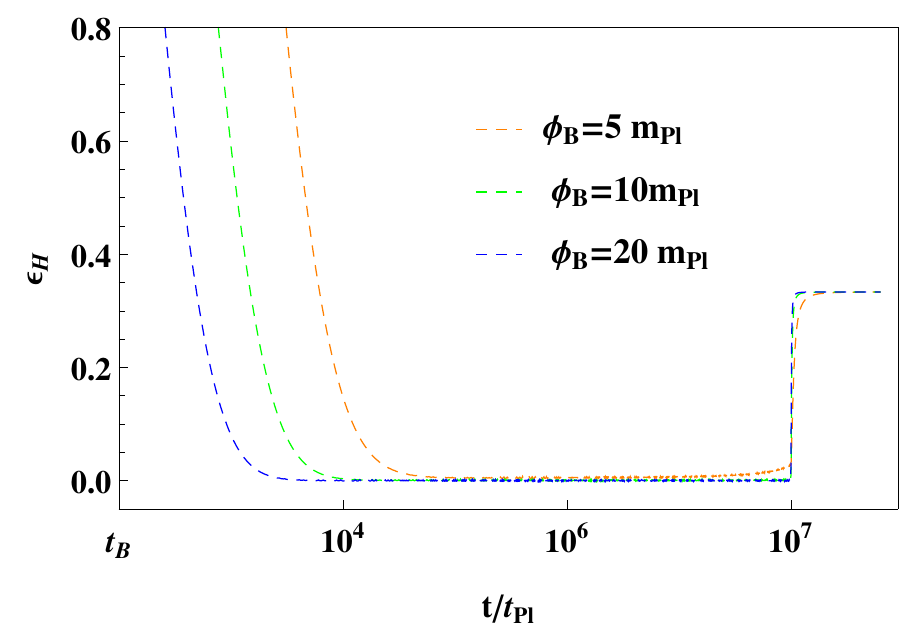}}
 \\
{\includegraphics[width=2.1in,height=1.6in,angle=0]{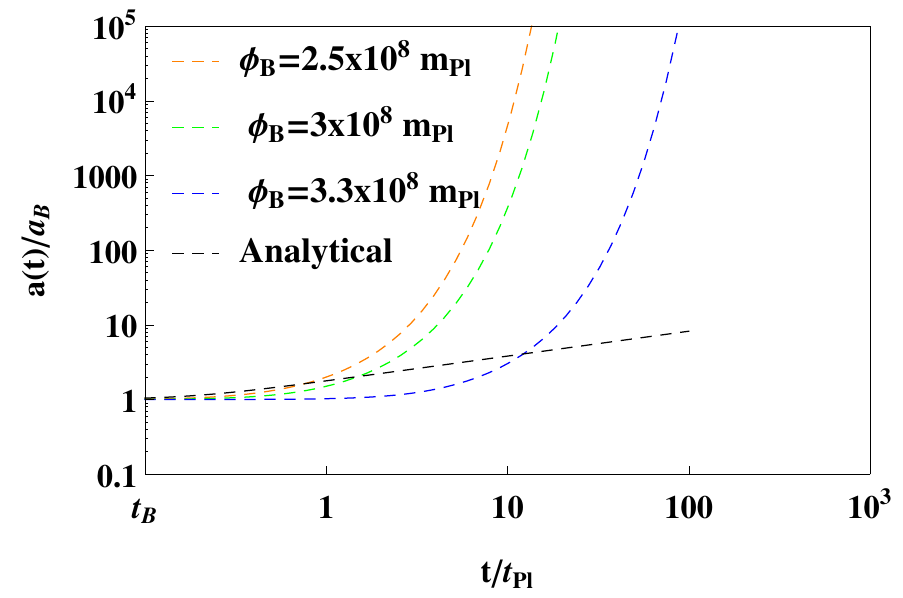}} & 
{\includegraphics[width=2.1in,height=1.6in,angle=0]{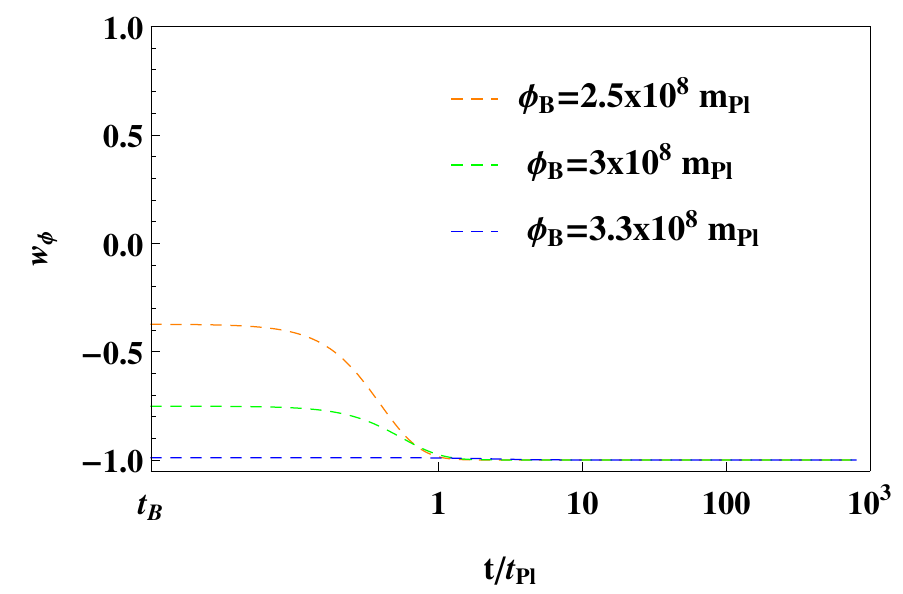}} & 
{\includegraphics[width=2.0in,height=1.6in,angle=0]{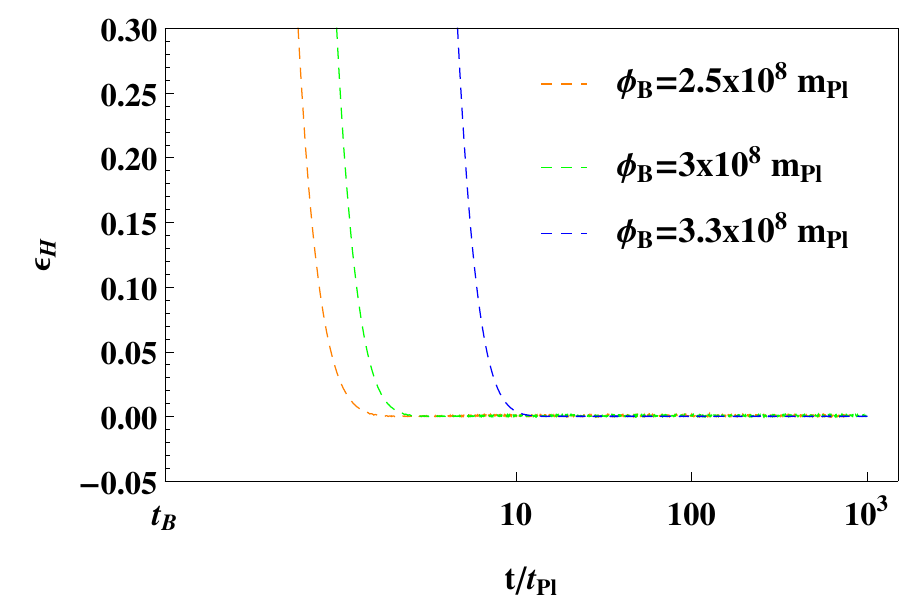}} 
\end{tabular}
\end{center}
\caption{ This figure represents the numerical solutions for the potential with $n=4/3$ and $\dot{\phi}_B<0$.}
\label{fig:4/3b}
\end{figure*}
\begin{figure*}[tbp]
\begin{center}
\begin{tabular}{ccc}
{\includegraphics[width=2.1in,height=1.6in,angle=0]{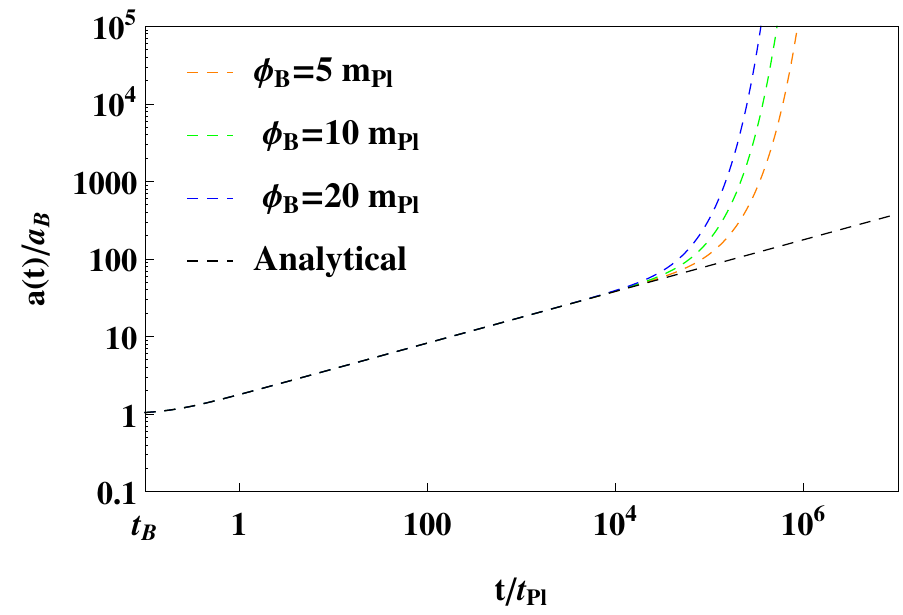}} &
{\includegraphics[width=2.1in,height=1.55in,angle=0]{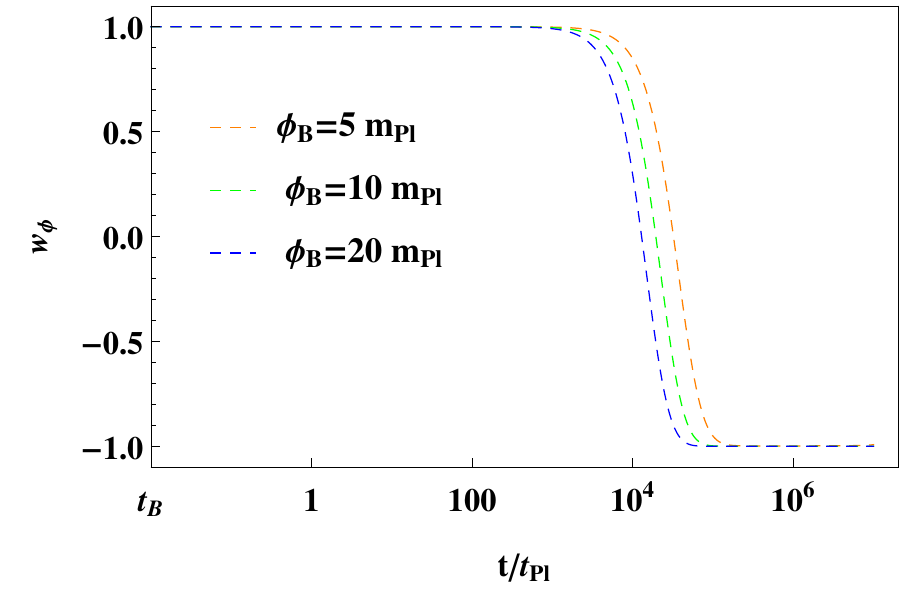}} &
{\includegraphics[width=2.0in,height=1.6in,angle=0]{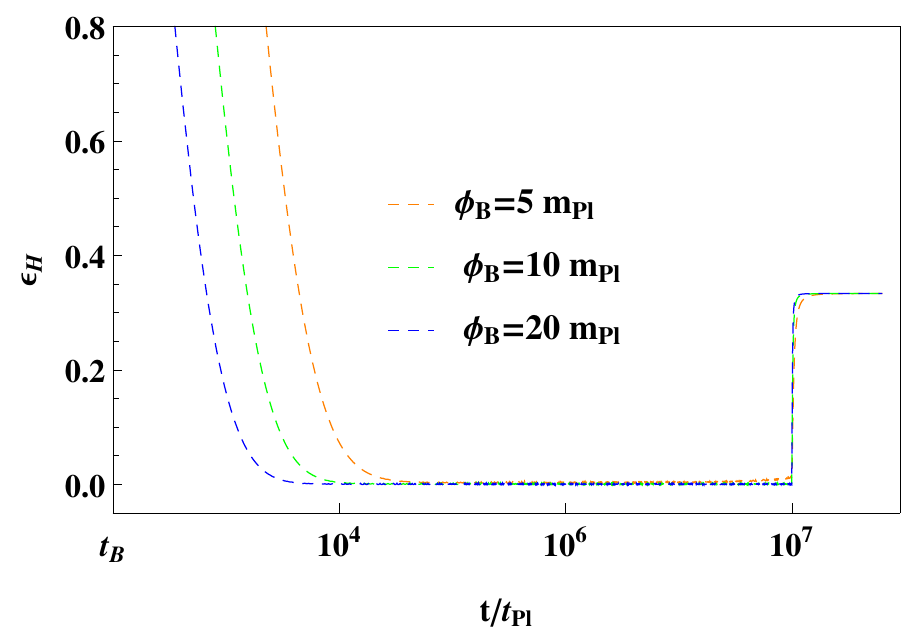}}
 \\
{\includegraphics[width=2.1in,height=1.6in,angle=0]{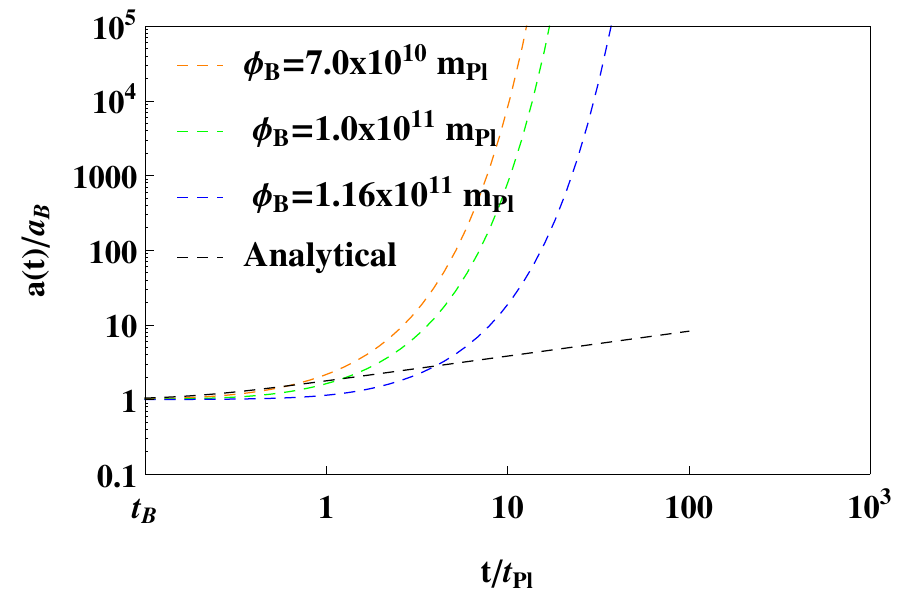}} & 
{\includegraphics[width=2.1in,height=1.6in,angle=0]{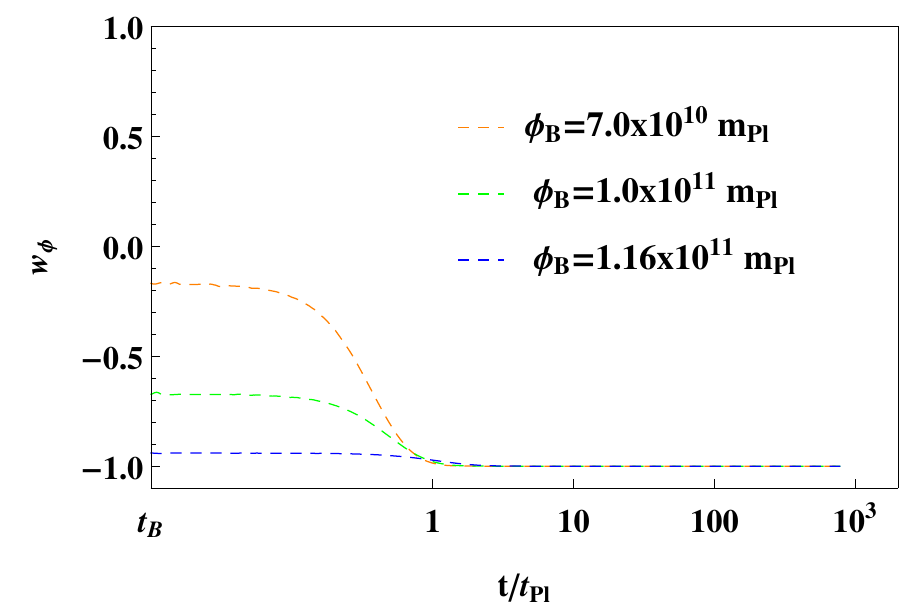}} & 
{\includegraphics[width=2.0in,height=1.6in,angle=0]{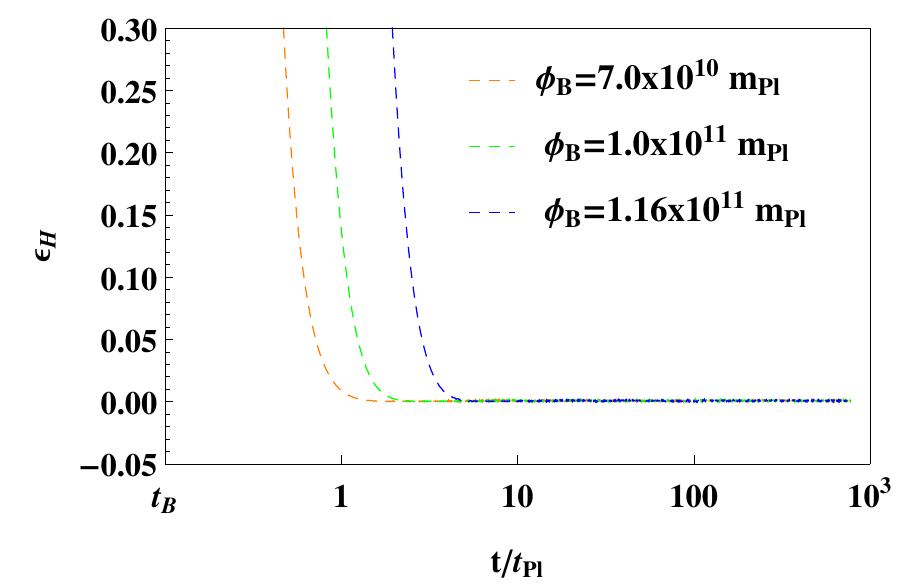}} 
\end{tabular}
\end{center}
\caption{ This figure is for $n=1$ and $\dot{\phi}_B<0$.}
\label{fig:n=1b}
\end{figure*}
\begin{figure*}[tbp]
\begin{center}
\begin{tabular}{ccc}
{\includegraphics[width=2.1in,height=1.6in,angle=0]{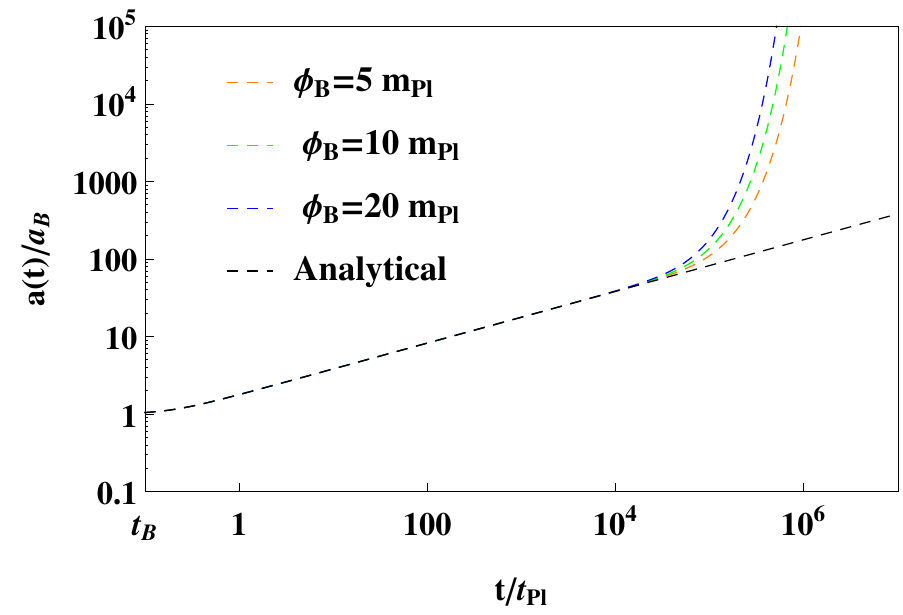}} &
{\includegraphics[width=2.1in,height=1.55in,angle=0]{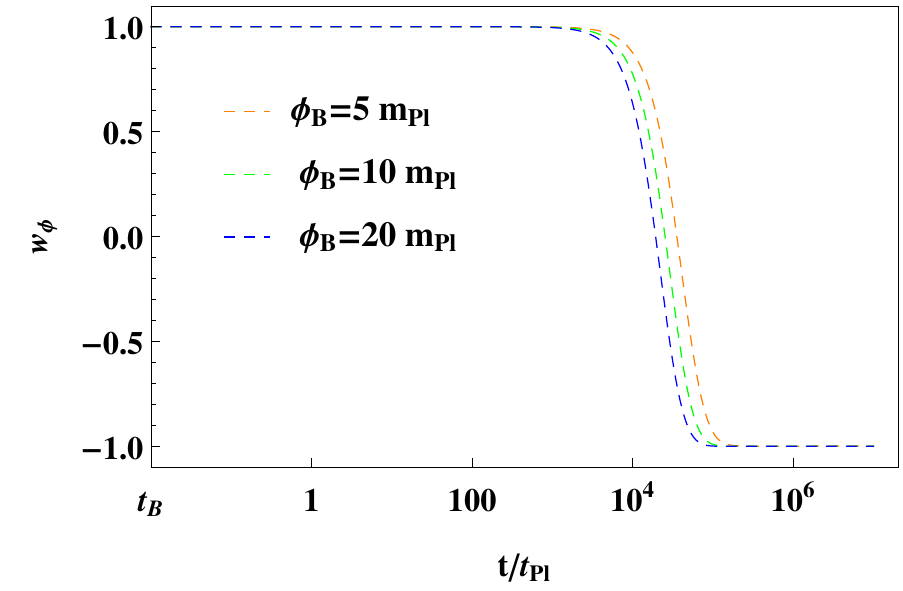}} &
{\includegraphics[width=2.0in,height=1.6in,angle=0]{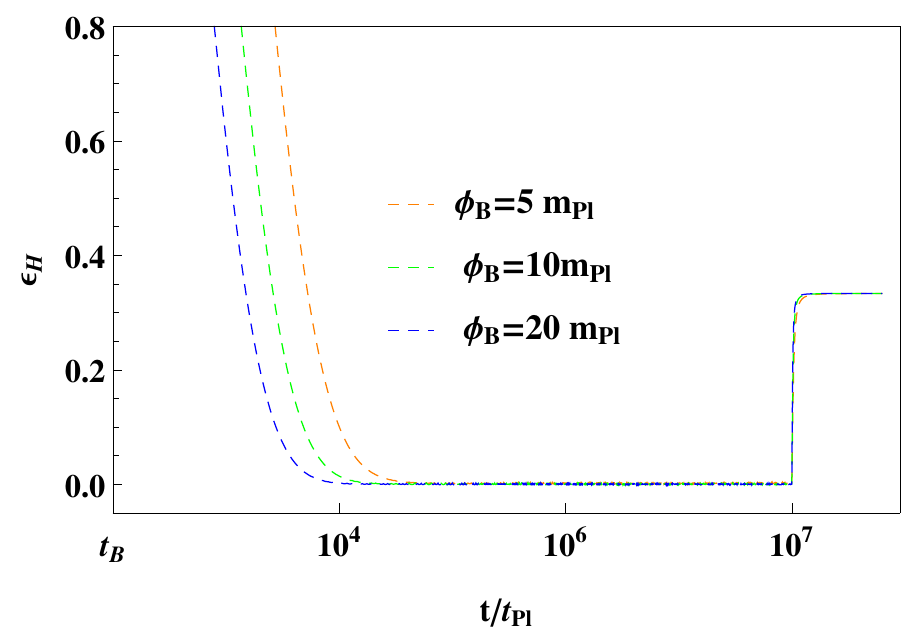}}
 \\
{\includegraphics[width=2.1in,height=1.6in,angle=0]{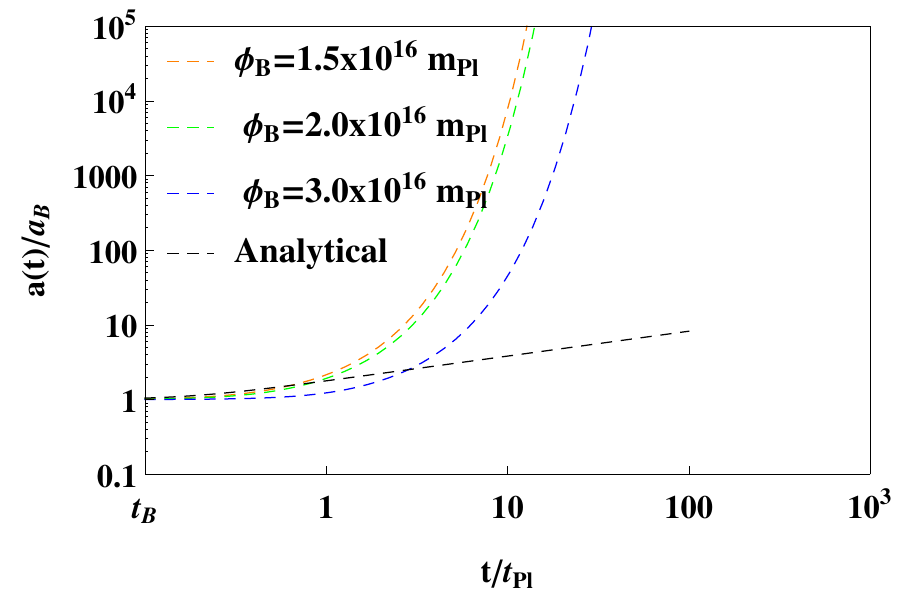}} & 
{\includegraphics[width=2.1in,height=1.6in,angle=0]{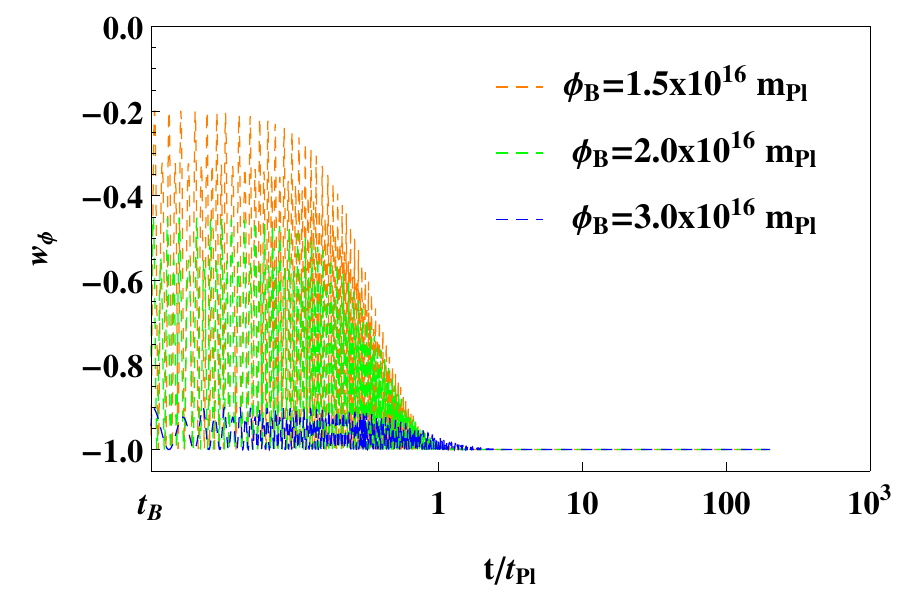}} & 
{\includegraphics[width=2.0in,height=1.6in,angle=0]{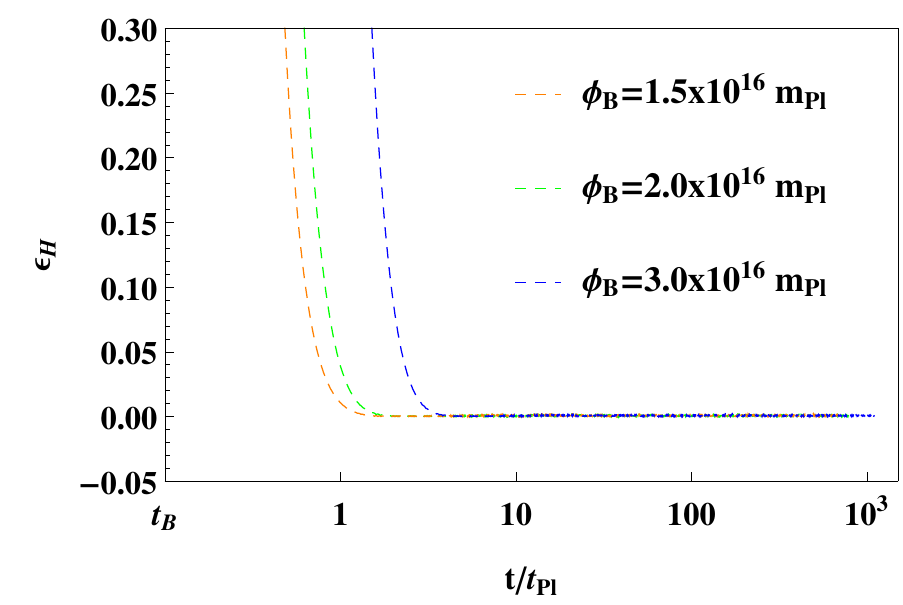}} 
\end{tabular}
\end{center}
\caption{ This figure corresponds to $n=2/3$ and $\dot{\phi}_B<0$.}
\label{fig:2/3b}
\end{figure*}
\begin{figure*}[tbp]
\begin{center}
\begin{tabular}{ccc}
{\includegraphics[width=2.1in,height=1.6in,angle=0]{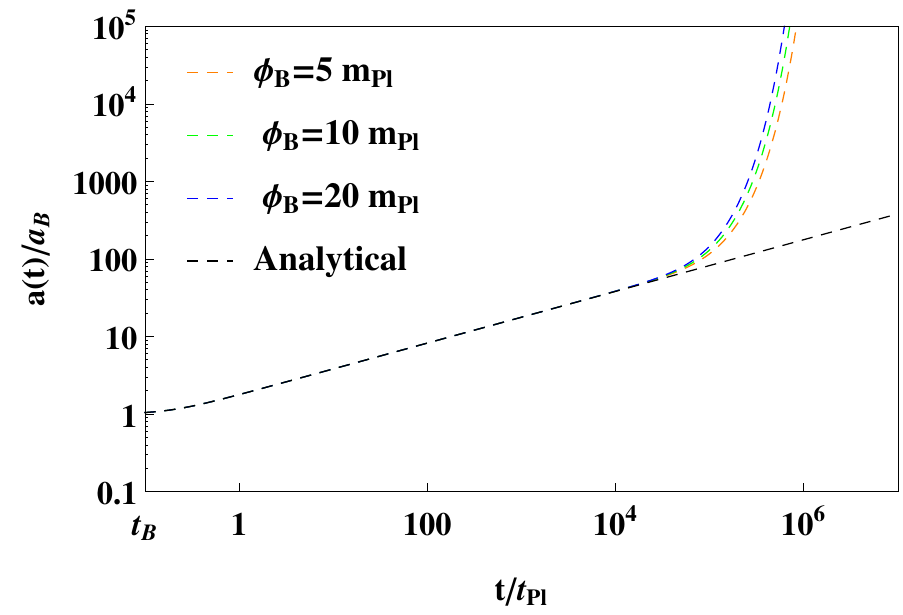}} &
{\includegraphics[width=2.1in,height=1.55in,angle=0]{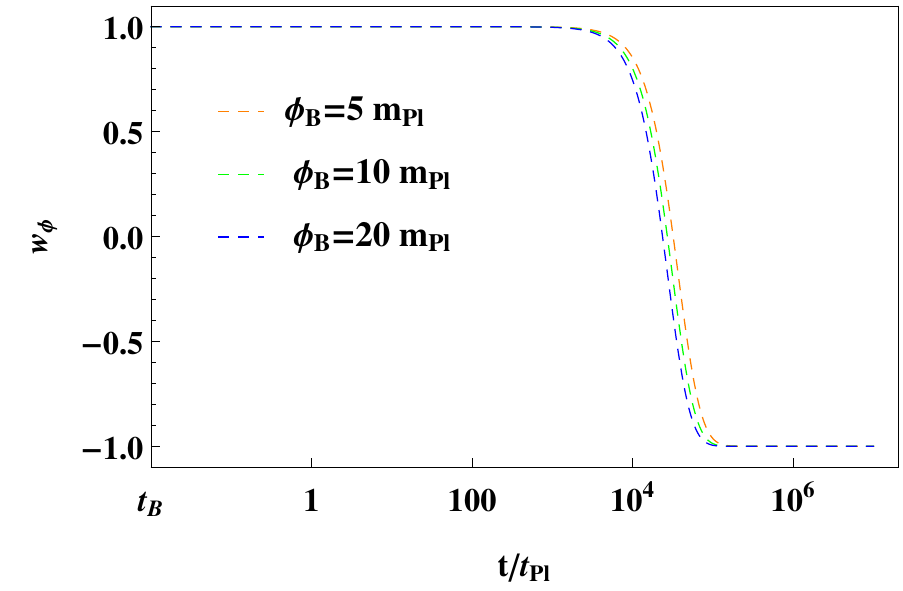}} &
{\includegraphics[width=2.0in,height=1.6in,angle=0]{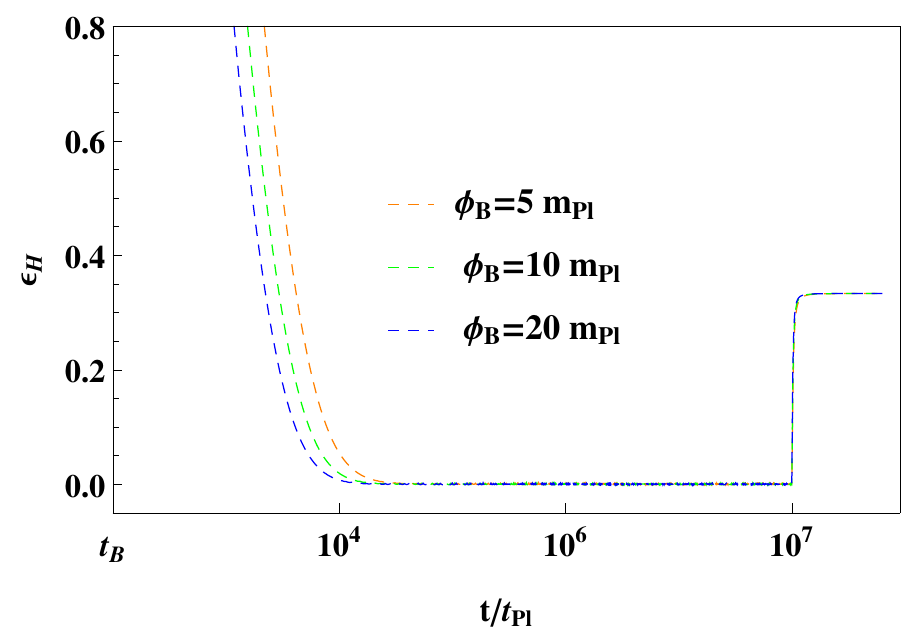}}
 \\
{\includegraphics[width=2.1in,height=1.6in,angle=0]{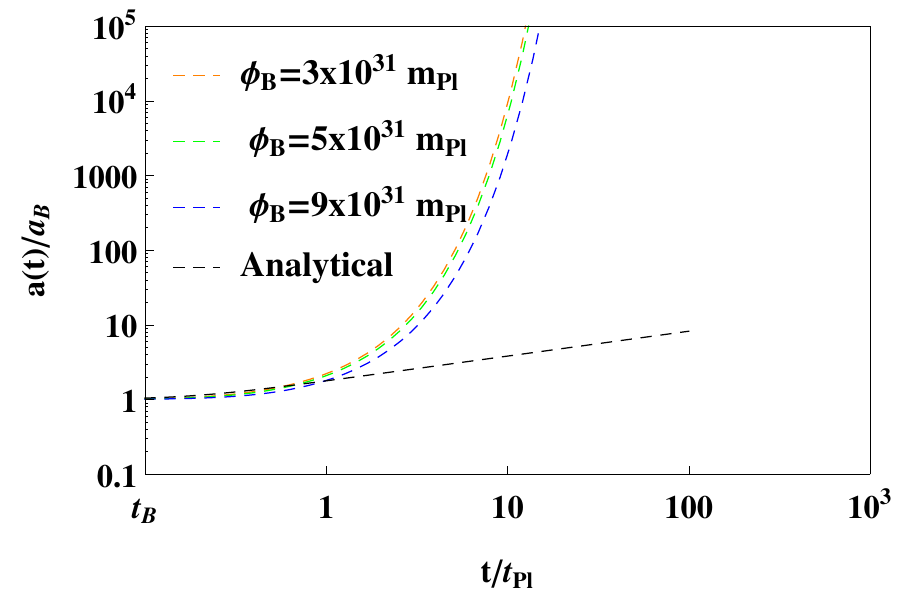}} & 
{\includegraphics[width=2.1in,height=1.61in,angle=0]{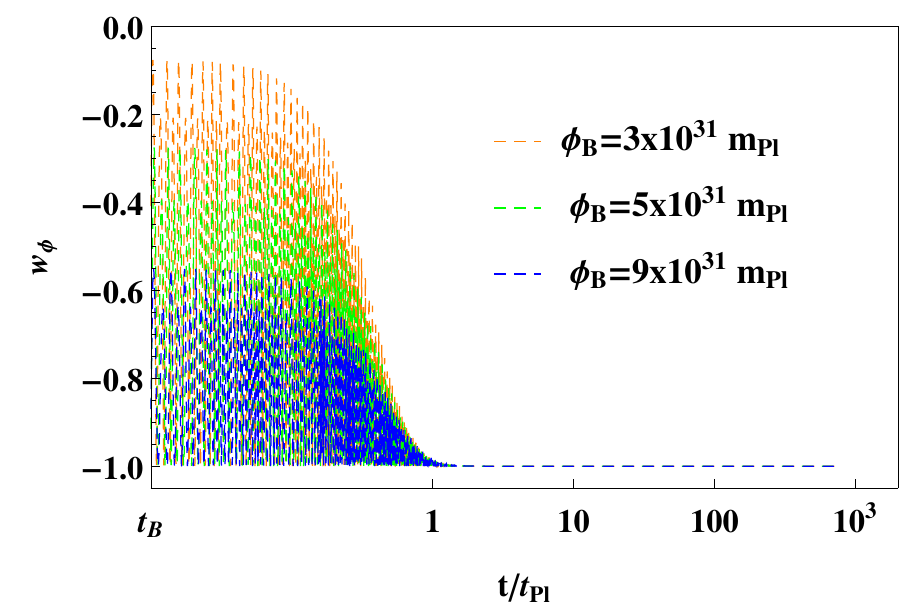}} & 
{\includegraphics[width=2.0in,height=1.56in,angle=0]{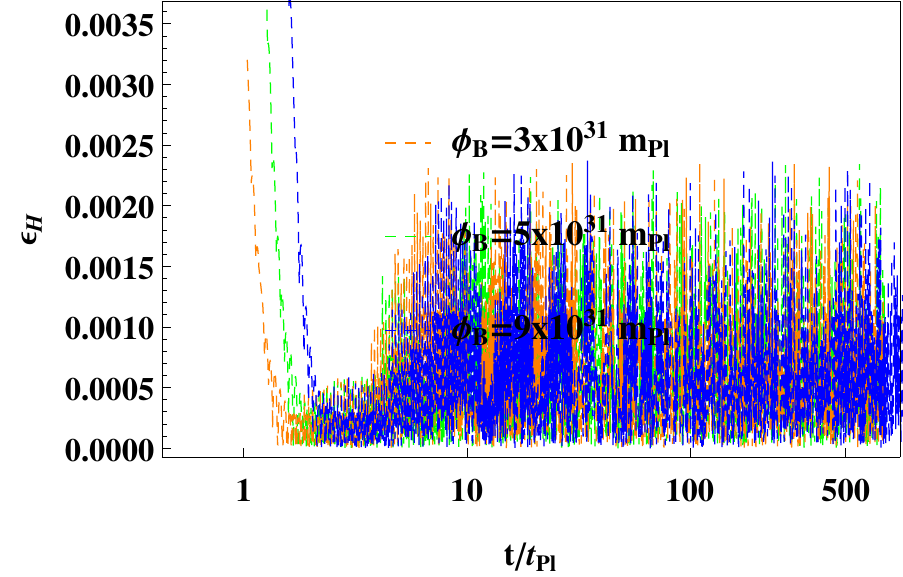}} 
\end{tabular}
\end{center}
\caption{ The numerical evolutions are displayed for $n=1/3$ and $\dot{\phi}_B<0$}
\label{fig:1/3b}
\end{figure*}
Finally, in the above cases, we consider the PED initial conditions at the bounce. The numerical results are illustrated in the bottom panels of Figs. \ref{fig:n=1a}, \ref{fig:2/3a} and \ref{fig:1/3a}, respectively. In each case, the evolution of $a(t)$ at the bounce is very sensitive to the set of initial conditions, and the universal features of it disappear. Specifically, the bouncing phase no longer exists. Although, the slow-roll regime $w(\phi)\simeq -1$ can still be achieved, see bottom panels of Figs. \ref{fig:n=1a}, \ref{fig:2/3a} and \ref{fig:1/3a}. Additionally, the PED initial conditions can lead to a large number of e-folds $N_{inf}$ during the slow-roll inflationary regime that are shown in Table \ref{tab:2/3a}.
\subsection{Negative inflaton velocity: $\dot{\phi}_B < 0$ }
\label{sec:phiB<0}
To cover the entire phase space we should also study the NIV at the bounce $(\dot{\phi}_B < 0)$. Similar to the previous sub-sections, in this sub-section, we choose the positive values of inflaton field at the bounce $({\phi}_B > 0)$ in order to keep the potential to be real. The detailed investigations for the power-law potential with $n<2$ are given below.
\subsubsection{Power-law potential with $n=7/4$}
\label{sec:n=7/4b}
Numerically, we evolve the system (\ref{eq:H}) and (\ref{eq:ddphi}) with (\ref{eq:pot}) and $n=7/4$. Again, we have two cases such as KED and PED at the quantum bounce. 

First we discuss the KED one, in which, top panels of Fig. \ref{fig:7/4b} show the evolution of $a(t)$, $w(\phi)$ and $\epsilon_H$ for a set of initial values of $\phi_B$. In the top left panel, we also exhibits the analytical solution of $a(t)$ (\ref{eq:a}) in comparison to the numerical solutions. Furthermore, we have three regimes; bouncing, transition and the slow-roll (Top middle panel of Fig. \ref{fig:7/4b}). Next, we calculate the number of e-folds $N_{inf}$ during the slow-roll regime for the restricted range of $\phi_B$, and given by (see Table \ref{tab:7/4b}) 
\begin{eqnarray}
\phi_B \in (4.3m_{pl}, \phi_{max})
\end{eqnarray}
where $\phi_{max}$ is given by Eq. (\ref{eq:phim7/4}). Note that we do not obtain the slow-roll regime with $\phi_B=0,0.5,1$ and 2 etc. for $(\dot{\phi}_B < 0)$. However, in sub-section \ref{sec:n=7/4a}, for the positive inflaton velocity $(\dot{\phi}_B > 0)$, the slow-roll regime covers the whole range of $\phi_B$ as $\phi_B \in (0, \phi_{max})$. To get at least 60 e-folds during the slow-roll inflation, $\phi_B$ has to be restricted as shown in Table \ref{tab:7/4b}:
\begin{eqnarray}
\phi_B \in (5.3m_{pl}, \phi_{max})
\end{eqnarray}
For the PED conditions, the evolution of  $a(t)$, $w(\phi)$ and $\epsilon_H$ are shown in the bottom panels of Fig. \ref{fig:7/4b} for different initial values of $\phi_B$. In this case, the universality of $a(t)$ has been lost and the bouncing phase does not exist any more. Though inflationary regime can be achieved. As $\phi_B$ grows, we get more number of e-folds, see Table \ref{tab:7/4b}.
\subsubsection{Power-law potential with $n=4/3, 1, 2/3$ and $1/3$}
\label{sec:n=4/3b}
Figs. \ref{fig:4/3b}, \ref{fig:n=1b}, \ref{fig:2/3b} and \ref{fig:1/3b} are plotted for $n=4/3, 1, 2/3$ and $1/3$, respectively, and show the evolutions of $a(t)$, $w(\phi)$ and $\epsilon_H$ for different sets of initial values of $\phi_B$. Top panels correspond to the KED case whereas bottom ones are for PED conditions. For each value of $n$, we get desired slow-roll inflationary phase. To obtain at least 60 e-folds during the slow-roll regime each $n$ has restricted range given as follows.\\
\\
For $n=4/3$,
\begin{eqnarray}
\phi_{B} \in (4.889m_{pl}, \phi_{max})
\end{eqnarray} 
for $n=1$,
\begin{eqnarray}
\phi_{B} \in (4.41m_{pl}, \phi_{max})
\end{eqnarray}
for $n=2/3$,
\begin{eqnarray}
\phi_{B} \in (4.02m_{pl}, \phi_{max})
\end{eqnarray}
and for $n=1/3$,
\begin{eqnarray}
\phi_{B} \in (3.49m_{pl}, \phi_{max})
\label{phim1/3}
\end{eqnarray}
where $\phi_{max}$ for $n=4/3, 1, 2/3$ and $1/3$ are given by the equations (\ref{eq:phim4/3}), (\ref{eq:phimn1}), (\ref{eq:phim2/3}) and (\ref{eq:phim1/3}), respectively.

The PED initial conditions lead to a large number of e-folds during the slow roll regime, and are shown in Tables \ref{tab:7/4b} and \ref{tab:2/3b}.
\begin{table*}
\caption{Table for $N_{inf}$, $r_{cw}(N_{inf} \simeq 60)$ and $r_w(60 < N_{inf} < 60)$ with $\dot{\phi}_B<0$. }
\begin{center}
\resizebox{\textwidth}{!}{%
\begin{tabular}{cccccccc}
\hline\hline
 $n$~~  & $m$ ~~& $\phi_B$~~~  & Inflation~~~ & $t/t_{pl}$~~~ & $\phi_{*}$~~~ & $N_{inf}$ &~~~$r_{cw}/r_w$\\
\hline\hline
\\
7/4 ~~ & $6.2 \times 10^{-6}$~~& 4.3~~~ & starts~~~& $7.18583 \times 10^4$ ~~~&2.10286~~~ & 29.4384 &~~~ $r_{cw} > r_w$\\
 ~~&~~&~~~& ends~~~& $9.531 \times 10^6$ ~~~& 0.169577~~~ &  \\\\
~~ &~~& 5~~~ & starts~~~& $5.48084 \times 10^4$ ~~~& 2.84702~~~ & 52.3782 &~~~ $r_{cw} > r_w$\\
 ~~&~~&~~~& ends~~~& $1.2252 \times 10^7$ ~~~& 0.378438~~~ &  \\\\ 
~~ &~~& 5.3~~~ & starts~~~& $4.99183 \times 10^4$ ~~~& 3.16226~~~ & 60.6001 &~~~ $r_{cw} = r_w$\\
 ~~&~~&~~~& ends~~~& $1.225 \times 10^7$ ~~~& 0.632925~~~ &  \\\\  
~~ &~~& 6.0~~~ & starts~~~& $4.15123 \times 10^4$ ~~~& 3.89232~~~ & 75.6016 &~~~ $r_{cw} < r_w$\\
~~&~~&~~~& ends~~~& $1.0707 \times 10^7$ ~~~& 2.25377~~~ &  \\\\    
~~ &~~& (PED)$4 \times 10^6$~~~ & starts~~~& 0.577 ~~~& $\simeq3.9 \times 10^6$~~~ & 713.8254 &~~~ $r_{cw} < r_w$\\
~~&~~&~~~& ends~~~& 1046.48 ~~~& $\simeq3.9 \times 10^6$~~~ &  \\ \\  
4/3 ~~ & $5.1 \times 10^{-5}$~~& 4.81~~~ & starts~~~& $5.19777 \times 10^4$ ~~~&2.66568~~~ & 57.2321 &~~~ $r_{cw} > r_w$\\
 ~~&~~&~~~& ends~~~& $1.0493 \times 10^7$ ~~~& 0.6593~~~ &  \\\\
 ~~ &~~& 4.889~~~ & starts~~~& $5.09254 \times 10^4$ ~~~& 2.74802~~~ & 60.1178 &~~~ $r_{cw} = r_w$\\
 ~~&~~&~~~& ends~~~& $1.0757 \times 10^7$ ~~~& 0.678072~~~ &  \\\\ 
 ~~ &~~& 5.0~~~ & starts~~~& $4.95217 \times 10^4$ ~~~& 2.86358~~~ & 62.1697 &~~~ $r_{cw} < r_w$\\
~~&~~&~~~& ends~~~& $1.0444 \times 10^7$ ~~~& 1.18179~~~ &  \\\\  
 ~~ &~~& 6.0~~~ & starts~~~& $4.0199 \times 10^4$ ~~~& 3.89758~~~ & 84.6534 &~~~ $r_{cw} < r_w$\\
~~&~~&~~~& ends~~~& $1.044 \times 10^7$ ~~~& 2.55716~~~ &  \\\\  
~~ &~~& (PED)$3 \times 10^8$~~~ & starts~~~& 0.665 ~~~& $\simeq3.0 \times 10^8$~~~ & 714.7624 &~~~ $r_{cw} < r_w$\\
~~&~~&~~~& ends~~~& $1178.2 $ ~~~& $\simeq3.0 \times 10^8$~~~ &  \\ \\  
1 ~~ & $1.9 \times 10^{-4}$~~& 4.34~~~ & starts~~~& $4.85251 \times 10^4$ ~~~&2.2069~~~ & 56.5358 &~~~ ${\color{blue}r_{cw} > r_w}$\\
 ~~&~~&~~~& ends~~~& $9.704 \times 10^6$ ~~~& 0.0783571~~~ &  \\\\
~~ &~~& 4.41~~~ & starts~~~& $4.77327 \times 10^4$ ~~~& 2.27959~~~ & 60.4555 &~~~ $r_{cw} = r_w$\\
 ~~&~~&~~~& ends~~~& $1.02023 \times 10^7$ ~~~& 0.0781092~~~ &  \\\\ 
  ~~ &~~& 4.45~~~ & starts~~~& $4.72885 \times 10^4$ ~~~& 2.32111~~~ & 62.7961 &~~~ $r_{cw} < r_w$\\
~~&~~&~~~& ends~~~& $1.0493 \times 10^7$ ~~~& 0.0772239~~~ &  \\\\  
~~ &~~& 5.0~~~ & starts~~~& $4.23001 \times 10^4$ ~~~& 2.88928~~~ & 80.3628 &~~~ $r_{cw} < r_w$\\
~~&~~&~~~& ends~~~& $1.0803 \times 10^7$ ~~~& 0.889376~~~ &  \\\\  
~~ &~~& (PED)$1.0 \times 10^{11}$~~~ & starts~~~& 0.575 ~~~& $\simeq1.0 \times 10^{11}$~~~ & 714.4728 &~~~ $r_{cw} < r_w$\\
~~&~~&~~~& ends~~~& $1048.3 $ ~~~& $\simeq1.0 \times 10^{11}$~~~ &  \\ \\  
\hline\hline
\end{tabular}}
\label{tab:7/4b}
\end{center}
\end{table*}
\begin{table*}
\caption{Table represents the number of e-foldings $N_{inf}$, $r_{cw}(N_{inf} \simeq 60)$ and $r_w(60 < N_{inf} < 60)$ with $\dot{\phi}_B<0$. }
\begin{center}
\resizebox{\textwidth}{!}{%
\begin{tabular}{cccccccc}
\hline\hline
 $n$~~  & $m$ ~~& $\phi_B$~~~  & Inflation~~~ & $t/t_{pl}$~~~ & $\phi_{*}$~~~ & $N_{inf}$ &~~~$r_{cw}/r_w$\\
\hline\hline
\\
2/3 ~~ & $4.7 \times 10^{-4}$~~& 3.0~~~ & starts~~~& $7.21753 \times 10^4$ ~~~&0.802212~~~ & 10.0481 &~~~ $r_{cw} > r_w$\\
 ~~&~~&~~~& ends~~~& $2.372 \times 10^6$ ~~~& 0.0418586 ~~~ &  \\\\
~~ & ~~& 4.0~~~ & starts~~~& $5.40072 \times 10^4$ ~~~&1.84949~~~ & 58.4348 &~~~ $r_{cw} > r_w$\\
 ~~&~~&~~~& ends~~~& $1.0068 \times 10^7$ ~~~& 0.0428617 ~~~ &  \\\\  
 ~~ &~~& 4.02~~~ & starts~~~& $5.37956 \times 10^4$ ~~~& 1.87013~~~ & 60.0514 &~~~ $r_{cw} = r_w$\\
 ~~&~~&~~~& ends~~~& $1.0349 \times 10^7$ ~~~& 0.00925105~~~ &  \\\\ 
 ~~ &~~& 4.2~~~ & starts~~~& $5.20966 \times 10^4$ ~~~& 2.05536 ~~~ & 67.2913 &~~~ $r_{cw} < r_w$\\
~~&~~&~~~& ends~~~& $1.0875 \times 10^7$ ~~~& 0.849753~~~ &  \\\\   
~~ &~~& 5.0~~~ & starts~~~& $4.6479 \times 10^4$ ~~~& 2.87395~~~ & 80.8130 &~~~ $r_{cw} < r_w$\\
~~&~~&~~~& ends~~~& $1.0772 \times 10^7$ ~~~& 1.47887~~~ &  \\\\  
~~ &~~& (PED)$3.0 \times 10^{16}$~~~ & starts~~~& 1.06 ~~~& $\simeq3.0 \times 10^{16}$~~~ & 716.1492 &~~~ $r_{cw} < r_w$\\
~~&~~&~~~& ends~~~& $1805.14 $ ~~~& $\simeq3.0 \times 10^{16}$~~~ &  \\ \\  
1/3 ~~ & $1.1 \times 10^{-3}$~~& 3.0~~~ & starts~~~& $5.13333 \times 10^4$ ~~~&0.857781~~~ & 22.3829 &~~~ $r_{cw} > r_w$\\
 ~~&~~&~~~& ends~~~& $3.368 \times 10^6$ ~~~& 0.0115291  ~~~ &  \\\\
~~ &~~& 3.49~~~ & starts~~~& $4.73607 \times 10^4$ ~~~& 1.3609~~~ & 60.5475 &~~~ $r_{cw} = r_w$\\
 ~~&~~&~~~& ends~~~& $8.3017 \times 10^6$ ~~~& 0.0114196~~~ &  \\\\  
~~ &~~& 4.0~~~ & starts~~~& $4.4815 \times 10^4$ ~~~& 1.87991 ~~~ & 82.7795 &~~~ $r_{cw} < r_w$\\
~~&~~&~~~& ends~~~& $1.0211 \times 10^7$ ~~~& 0.872017~~~ &  \\\\
~~ &~~& 5.0~~~ & starts~~~& $4.16413 \times 10^4$ ~~~& 2.89187 ~~~ & 93.8514 &~~~ $r_{cw} < r_w$\\
~~&~~&~~~& ends~~~& $1.067 \times 10^7$ ~~~& 2.71606~~~ &  \\\\   
~~ &~~& (PED)$3.0 \times 10^{31}$~~~ & starts~~~& 0.3109 ~~~& $\simeq3.0 \times 10^{31}$~~~ & 715.3383 &~~~ $r_{cw} < r_w$\\
~~&~~&~~~& ends~~~& $781.16 $ ~~~& $\simeq3.0 \times 10^{31}$~~~ &  \\ \\ 
\hline\hline
\end{tabular}}
\label{tab:2/3b}
\end{center}
\end{table*}
\begin{figure}[tbp]
\begin{center}
\begin{tabular}{c}
{\includegraphics[width=2.3in,height=2.2in,angle=0]{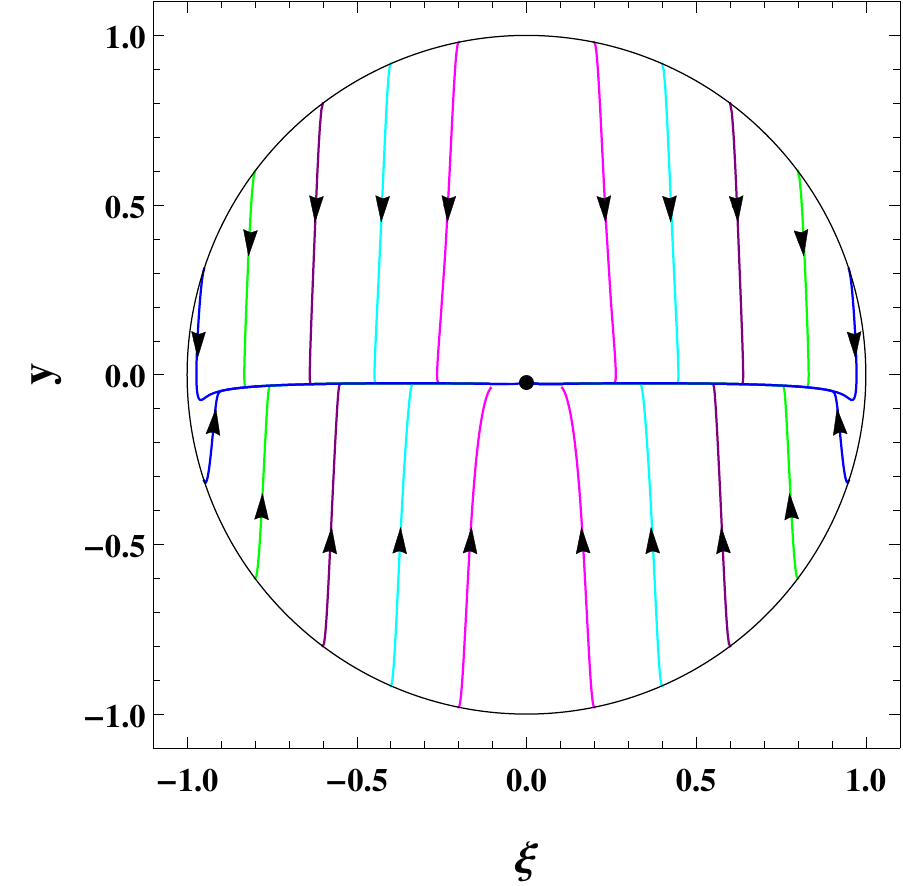}} \\
\\
{\includegraphics[width=2.3in,height=2.2in,angle=0]{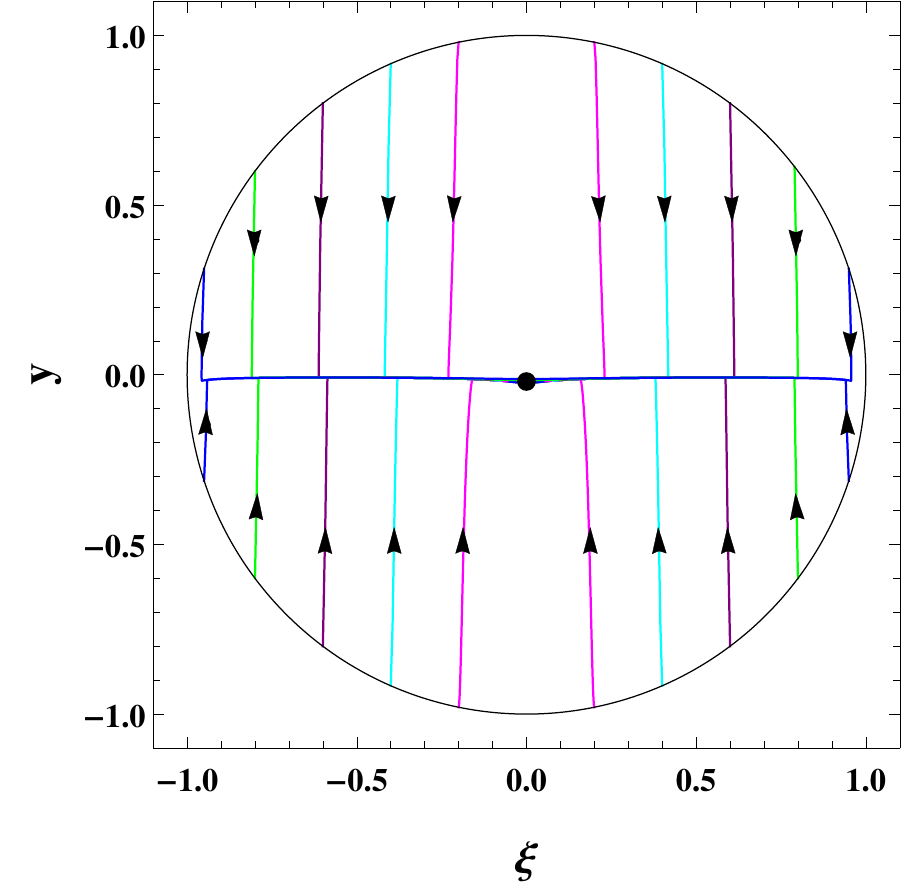}} 
\\
{\includegraphics[width=2.3in,height=2.2in,angle=0]{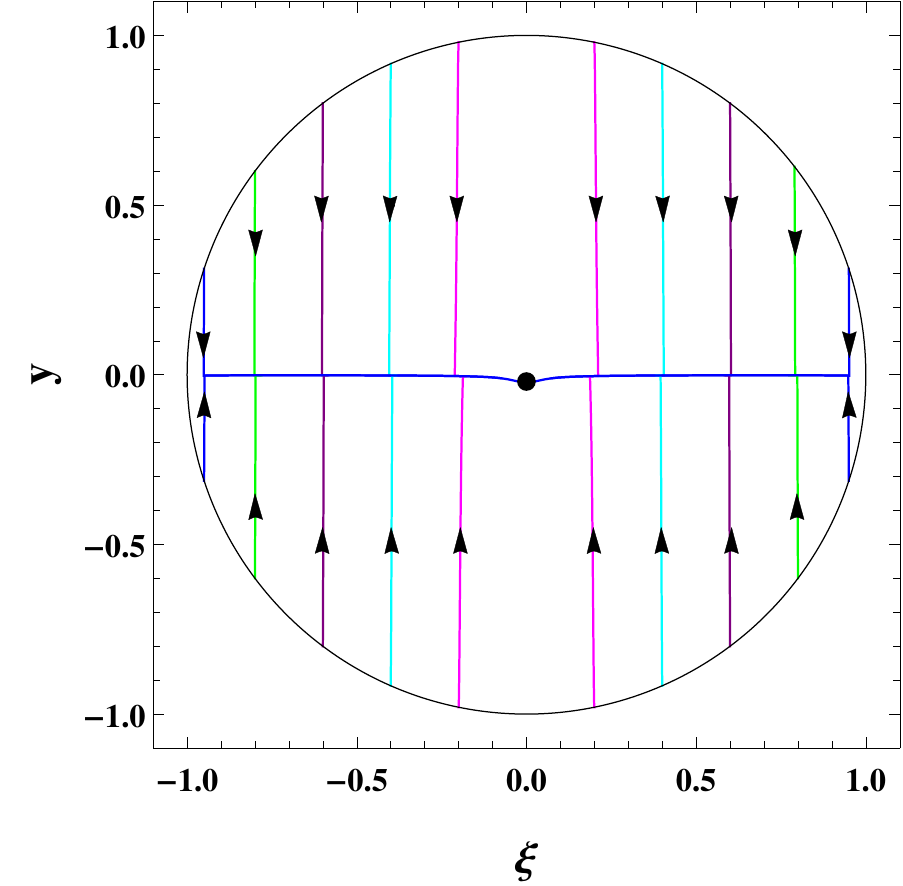}} 
\\
\end{tabular}
\end{center}
\caption{This figure shows the evolution of phase space trajectories for the power-law potential with $n=7/4$ (Top), $4/3$ (Middle) and 1 (Bottom). The arrows represent the direction of the trajectories from boundary to the origin. For the better depiction, we choose $m=0.2 m_{pl}$.}
\label{fig:port1}
\end{figure}
\begin{figure}[tbp]
\begin{center}
\begin{tabular}{c}
{\includegraphics[width=2.3in,height=2.2in,angle=0]{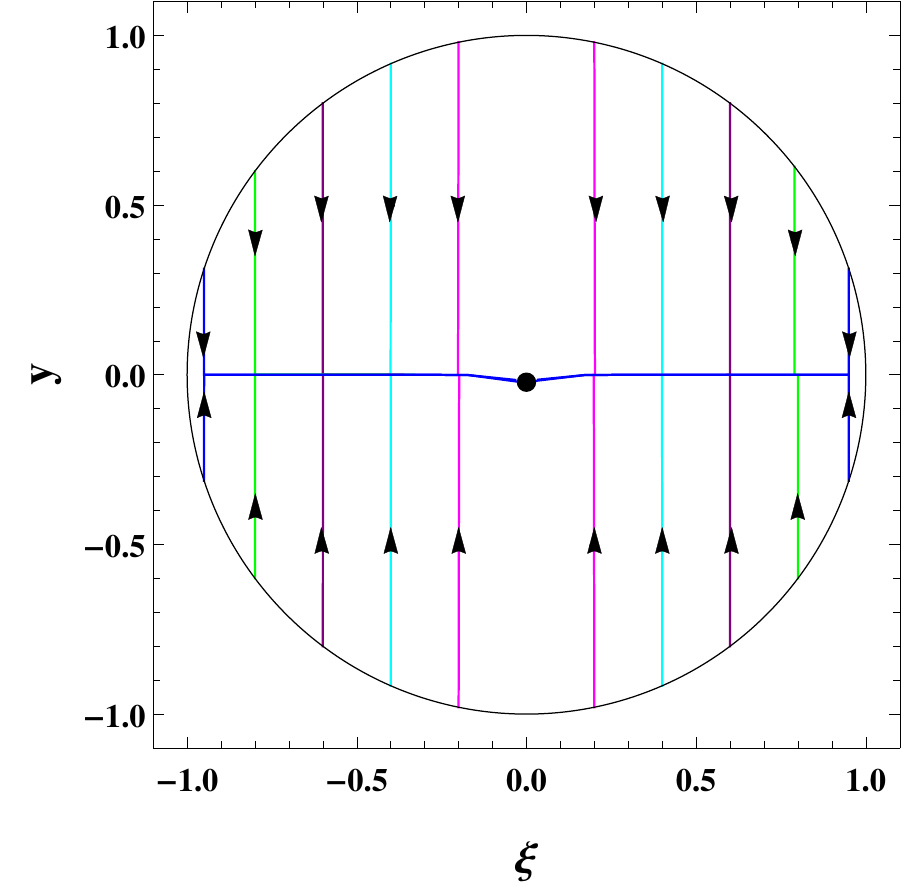}} 
\\
{\includegraphics[width=2.3in,height=2.2in,angle=0]{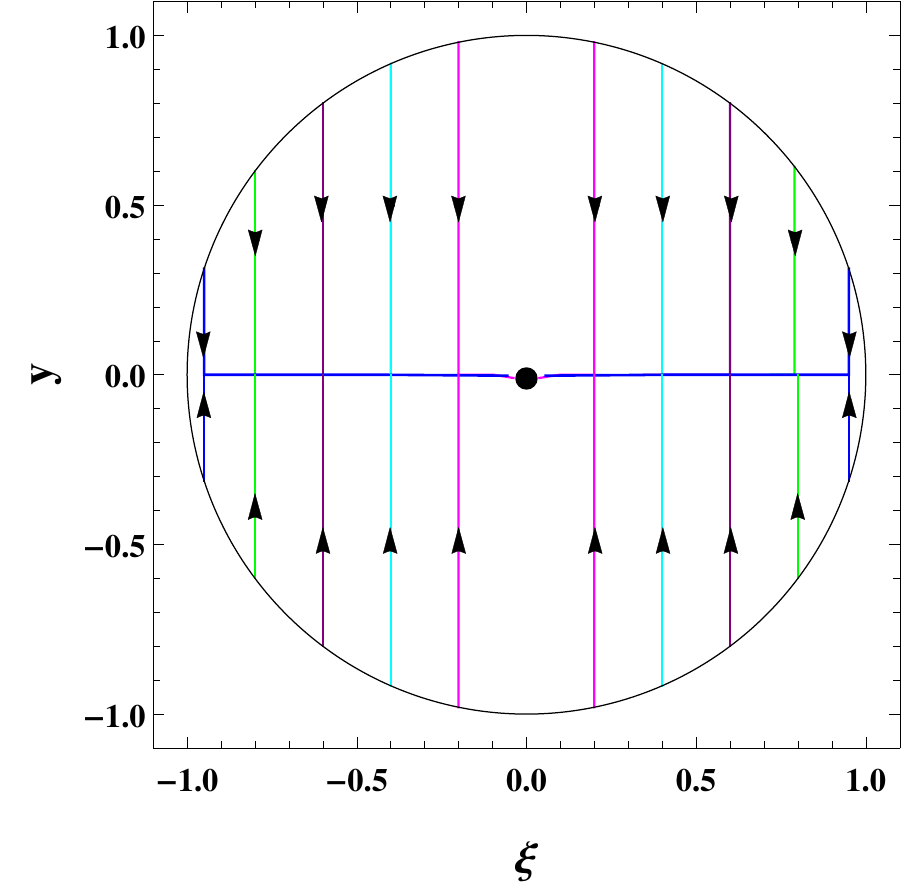}} 
\end{tabular}
\end{center}
\caption{This figure corresponds to the potential with $n=2/3$ (Top) and $1/3$ (Bottom). The arrows show the time evolution from boundary to the origin. We use $m=0.2 m_{pl}$.}
\label{fig:port2}
\end{figure}
\section{Phase portrait and the desired slow-roll inflation}
\label{sec:phase}
According to Planck 2015 results \cite{Planck2015}, the power-law potential with $n=2$ is moderately disfavored compared to the models predicting a smaller tensor-to-scalar ratio, such as $R^2$ inflationary model proposed by Starobinsky \cite{staro1980}. However, the power-law potential with $n<2$ is consistent with the Planck 2015 results. Therefore, in this section, we shall study the phase space analysis for the power-law potentials with $n=7/4, 4/3, 1, 2/3$ and $1/3$, respectively. The detailed analysis are given below.

As we mentioned in section \ref{sec:phiB>0}, the inflaton field $\phi$ must be positive, to get the real potential. Conclusively, it would cover only the half phase space (semi-circle). To obtain the whole phase space, more precisely, to show the phase space trajectories in a complete circle, we introduce the following dimensionless quantities\\
\begin{eqnarray}
\xi=\pm \sqrt{\frac{V(\phi)}{\rho_c}}, \qquad y=\pm \frac{\dot{\phi}}{\sqrt{2 \rho_c}}, \qquad \tau=m t
\label{eq:dimquant}
\end{eqnarray}
to form an autonomous system of evolution equations, which are given by
\begin{align}
\frac{d \xi}{d\tau}&= \pm \sqrt{2V_0} \frac{n}{2m} \left( \frac{\rho_c \xi^2}{V_0} \right)^{\frac{n-2}{2n}} y \nonumber\\
 \frac{d y}{d\tau}&= -3Hy/m -  \frac{V_0 n}{m\sqrt{2\rho_c}} \left( \frac{\rho_c \xi^2}{V_0} \right)^{\frac{n-1}{n}}
\label{eq:auto}
\end{align}
where $V(\phi)=V_{0} \phi^n$, and $V_0=1/2~ m^{4-n}$. At the bounce, we have $\rho=\rho_c$, this implies that
\begin{eqnarray}
\xi_B^2 + y_B^2 =1
\label{eq:circle}
\end{eqnarray}
This is the equation  of a circle. Thus, for the chosen dimensionless variables, we obtain equation of the circle at the quantum bounce irrespective of $n$.

First we do analysis for $n=7/4$. Tables \ref{tab:7/4a} and \ref{tab:7/4b} are obtained for PIV $(\dot{\phi}_B>0)$ and NIV $(\dot{\phi}_B<0)$. By looking at both tables, we observe that the observationally compatible initial conditions are $\phi_B \geq 0.805 m_{pl}$ for $\dot{\phi}_B>0$ and $\phi_B \geq 5.3 m_{pl}$ for $\dot{\phi}_B<0$. Therefore, in the whole parameter space of initial conditions these are only the KED initial conditions which can produce the desired slow-roll inflationary regime in their future evolution. However, some of the KED initial conditions do not provide the desired slow-roll, see the values of $\phi_B$ in Tables \ref{tab:7/4a} and \ref{tab:7/4b}. All the PED initial conditions at the bounce can lead to the desired slow-roll phase, and a large number of e-folds can be obtained, as shown in Tables \ref{tab:7/4a} and \ref{tab:7/4b}.

Top panel of Fig. \ref{fig:port1} shows the evolution of the phase space trajectories for both PIV and NIV, and also for both the KED and PED initial conditions, more accurately, it corresponds to the entire phase space. Regions close to the boundary of the circle represent the large energy density with the dominance of the quantum effects whereas small energy limit exists near the origin in the $\xi-y$ plane. All the phase space trajectories are started from the bounce $(\rho=\rho_c)$, and directed towards the origin which is the only stable point. The main characteristic of these portraits is the inflationary separatrix \cite{psingh2006} that is shown in the figure where all the trajectories are attracted towards the origin.

Second, we consider the power-law potential with $n=4/3$. In this case, we obtain 60 or more e-folds at the initial conditions $\phi_B \geq 0.4005 m_{pl}$ for $\dot{\phi}_B>0$ and $\phi_B \geq 4.889 m_{pl}$ for $\dot{\phi}_B<0$. Similar to the case of $n=7/4$, here also some KED initial conditions do not provide the desired slow-roll phase. In the case of PED initial conditions, the enormous amount of inflation is obtained, see Tables \ref{tab:7/4a} and \ref{tab:7/4b}. The phase portrait for this case is exhibited in the middle panel of Fig. \ref{fig:port1}.

Third, we study the potential with $n=1$. For PIV  $(\dot{\phi}_B>0)$, the whole range of KED initial conditions of $\phi_B$ $(\phi_B \geq 0)$ is consistent with the observations whereas in case of NIV $(\dot{\phi}_B<0)$, it should be restricted as  $\phi_B \geq 4.41 m_{pl}$. This implies that, in the case of $\dot{\phi}_B>0$, the entire range of KED initial conditions $(\phi_B \geq 0)$ can produce the desired slow-roll phase, and lead to more than 60 e-folds which do not possible in the cases of $n=7/4$ and $4/3$. All the PED initial conditions can give rise to a large amount of inflation, see Tables \ref{tab:2/3a} and \ref{tab:7/4b}. The corresponding phase portrait is shown in the bottom panel of Fig. \ref{fig:port1}.

Next, we focus on the potential with $n=2/3$. Here also, for $\dot{\phi}_B>0$, the entire range of $\phi_B$ $(\phi_B \geq 0)$ is in a good agreement with the observations whereas for $\dot{\phi}_B<0$, it is restricted $\phi_B \geq 4.02 m_{pl}$ that is shown in Tables \ref{tab:2/3a} and \ref{tab:2/3b}. Top panel of Fig. \ref{fig:port2} demonstrates the phase space trajectories for the case under consideration.

Finally, we deal with the potential $n=1/3$. In this case, the observationally compatible KED initial conditions are $\phi_B \geq 0$ for $\dot{\phi}_B>0$ and $\phi_B \geq 3.49 m_{pl}$ for $\dot{\phi}_B<0$. The PED initial conditions provide a large number of e-folds, see Tables \ref{tab:2/3a} and \ref{tab:2/3b}. The phase portrait in this case is shown in the bottom panel of Fig. \ref{fig:port2}.

We are now in position to compare our results with the consequences that are studied in the literature for the quadratic and Starobinsky potentials \cite{chen2015,Tao2017,Bonga2016}. Our studies for the power-law potentials with $n<2$ are in compatible with the quadratic potential. However, quadratic potential has been almost ruled out by the Planck data \cite{Planck2015}. In terms of number of e-folds, both KED and PED initial conditions are in good agreement with observations for the potentials with $n<2$ while Starobinsky inflation is observationally consistent only for KED initial conditions and not for PED ones at the bounce \cite{Tao2017,Bonga2016}.
\section{Conclusions}
\label{sec:conc}

In the framework of LQC, we studied the pre-inflationary dynamics of the power-law potential with $n<2$ for PIV and NIV, and also for KED and PED cases. We restricted ourselves to the said potential as the quadratic potential has been moderately disfavored by the Planck 2015 data \cite{Planck2015}. In addition, we chose only positive values of inflaton field in order to keep the potential to be real. First, we considered PIV at the quantum bounce, the evolution of background is divided into KED and PED cases. In case of KED initial conditions, the universe is always split into three different phases prior to the preheating, {\em bouncing, transition and the slow-roll}. In the bouncing phase, the background evolution is independent not only on the wide varieties of initial conditions but also the potentials. Specifically, the numerical evolution of the scale factor has shown the universal feature and compared by the analytical solution (\ref{eq:a}), see upper panels of Figs. \ref{fig:7/4a}$-$\ref{fig:1/3a}. During this phase, the EOS remains practically $w(\phi) \simeq 1$. However, in transition phase, it decreases rapidly from $w(\phi) \simeq 1$ to $w(\phi) \simeq -1$. The period of the transition regime is very short in comparison with other two regimes. Thereafter, the universe enters into an accelerating regime, where the slow-roll parameter $\epsilon_H$ is still large, later it drastically decreases to almost zero, by which the slow-roll inflation commences, as shown in upper panels of Figs. \ref{fig:7/4a}$-$\ref{fig:1/3a}. Next, we calculated the number of e-folds $N_{inf}$ during the slow-roll inflation, and presented in Tables \ref{tab:7/4a} and \ref{tab:2/3a}. In case of PED initial conditions, the universality of expansion factor $a(t)$ disappears (see bottom panels of Figs. \ref{fig:7/4a}$-$\ref{fig:1/3a}). One can get the slow-roll inflation for a long period, and correspondingly a large number of e-folds $N_{inf}$ are obtained, as displayed in bottom panels of Figs. \ref{fig:7/4a}$-$\ref{fig:1/3a} and Tables \ref{tab:7/4a} and \ref{tab:2/3a}.

Second, we considered NIV at the bounce, in this case also the evolution for KED initial conditions is divided into three regimes, namely bouncing, transition and the slow-roll. The universal feature of expansion factor appears whereas it is lost in case of PED initial conditions. The evolution of $a(t)$, $w(\phi)$ and $\epsilon_H$ are shown in Figs. \ref{fig:7/4b}$-$\ref{fig:1/3b}, and corresponding number of e-folds are presented in Tables \ref{tab:7/4b} and \ref{tab:2/3b}. To be consistent with the current observations at least 60 e-folds are needed during the slow-roll inflation. In case of PIV $(\dot{\phi}_B>0)$, to get at least 60 or more e-folds, we have restricted the range of $\phi_B$ for $n=7/4$ and $4/3$ whereas the restriction has  disappeared for $n=1, 2/3$ and $1/3$, see Tables \ref{tab:7/4a} and \ref{tab:2/3a}. In contrast, in case of NIV $(\dot{\phi}_B<0)$, all cases of $n$ have restricted range, as shown in Tables \ref{tab:7/4b} and \ref{tab:2/3b}.

We also presented phase space analysis for the inflationary potentials under consideration. As we mentioned, the inflaton field $\phi$ must be positive in order for the potential to be real. But, it corresponds to only half phase space. Therefore, we used a generic set of dynamical variables to bring out better depiction of underlying dynamics. The phase portraits for the power-law potentials with $n=7/4, 4/3, 1, 2/3$ and $1/3$ are exhibited in Figs. \ref{fig:port1} and \ref{fig:port2}. Moreover, for the generic initial conditions, the slow-roll inflation is an attractor in the phase space.
\section*{Acknowledgements}
We would like to thank J.D. Barrow and  T. Zhu for valuable  discussions and suggestions. A.W. is supported in part by the National Natural Science Foundation of China (NNSFC) with the Grants Nos. 11375153 and 11675145.

\end{document}